\documentclass[manuscript]{aastex}

\slugcomment{Submitted version November 23, 2007}

\begin{document}


\title{Identification of RR Lyrae Variables in SDSS from Single-Epoch 
Photometric and Spectroscopic Observations}

\author{Ronald Wilhelm\altaffilmark{1,7}, 
 W. Lee Powell, Jr.\altaffilmark{1,7}, Timothy C. Beers\altaffilmark{2},
Branimir Sesar\altaffilmark{3}, Carlos Alende Prieto\altaffilmark{4}, 
Kenneth W. Carrell\altaffilmark{1,7},Young Sun Lee\altaffilmark{2},
Brian Yanny\altaffilmark{5}, Constance M. Rockosi\altaffilmark{6},
Nathan De Lee\altaffilmark{2},Gwen Hansford Armstrong\altaffilmark{1,7}, 
Stephen J. Torrence\altaffilmark{1,7}}

\altaffiltext{1}{Physics Department, Texas Tech University,
    Lubbock, TX 79409, email: ron.wilhelm@ttu.edu, kenneth.w.carrell@ttu.edu, 
  william.l.powell@ttu.edu}

\altaffiltext{2}{Department of Physics and Astronomy, CSCE: Center for
  the Study of Cosmic Evolution, and JINA: Joint Institute for Nuclear
  Astrophysics, Michigan State University, East Lansing, MI
  48824,  email: beers@pa.msu.edu, leeyou25@msu.edu,
delee@pa.msu.edu} 

\altaffiltext{3}{Department of Astronomy, University of Washington, Box 351580,
Seattle, WA 98195-1580, email: 
bsesar@astro.washington.edu}

\altaffiltext{4}{McDonald Observatory and the Department of Astronomy, University of Texas,
Austin, TX 78712, email: callende@astro.as.utexas.edu}

\altaffiltext{5}{Fermi National Accelerator Laboratory, Box 500, Batavia, IL
60510, email: yanny@fnal.edu}

\altaffiltext{6}{University of California-Santa Cruz, 1156 High St., Santa Cruz,
CA 95060, email: crockosi@ucolick.org} 

\altaffiltext{7}{Visiting Astronomer, McDonald Observatory}




\begin{abstract}

We describe a new RR Lyrae identification technique based on out-of-phase
single-epoch photometric and spectroscopic observations contained in 
SDSS Data Release 6 (DR-6). This technique detects variability by exploiting the large
disparity between the $g-r$ color and the strength of the hydrogen Balmer lines
when the two observations are made at random phases. Comparison with a large sample
of known variables in the SDSS equatorial stripe (Stripe 82) shows that the
discovery efficiency for our technique is $\sim85\%$. Analysis of stars with
mulitiple spectroscopic observations suggests a similar efficiency throughout
the entire DR-6 sample. We also develop a technique to estimate the
average $g$ apparent magnitude (over the pulsation cycle) for individual RR
Lyrae stars, using the $<g-r>$ for the entire
sample and measured colors for each star. The resulting distances are found
to have precisions of $\sim \pm14\%$. Finally, we explore the properties of our
DR-6 sample of $N = 1087$ variables, and recover portions of the Sagittarius
Northern and Southern Stream. Analysis of the distance and velocity for the
Southern Stream are consistent with previously published data for blue
horizontal-branch stars. In a sample near the North Galactic Polar Cap, we find
evidence for the descending leading Northern arm, and a possible detection of
the trailing arm.

\end{abstract}





\section{Introduction}

RR Lyrae variable stars are extremely useful probes of the Galactic halo. Their
unique light curves allow for confident identification, while their intrinsic
high luminosities allows them to be observed to the outer limits of the Milky
Way. Although RR Lyrae variables have been used in the past to explore the local
Galactic halo (e.g., Layden 1994)), studies of the outer halo with similarly
large samples have only just begun. These new explorations have been made
possible by the enormous amounts of photometric data from surveys such as QUEST
(Vivas \& Zinn 2006) and the Sloan Digital Sky Survey (SDSS; York et al. 2000).

The Sloan Digital Sky Survey uses a CCD camera (Gunn et al. 1998) on a dedicated
2.5m telescope (Gunn et al. 2006) at Apache Point Observatory, New Mexico, to
obtain images in five broad optical bands ($ugriz$; Fukugita et al.~1996) over
approximately 10,000~deg$^2$ of the high Galactic latitude sky. The survey
data-processing software measures the properties of each detected object in the
imaging data in all five bands, and determines and applies both astrometric and
photometric calibrations ( Lupton et al. 2001; Pier et al. 2003; Ivezi\'c et
al.~2004). Photometric calibration is provided by simultaneous observations with
a 20-inch telescope at the same site (Hogg et al.~2001; Smith et al.~2002;
Stoughton et al.~2002; Tucker et al.~2006).

The SDSS and the Sloan Extension for Galactic Understanding and Exploration
(SEGUE) have now imaged over 9,500 square degrees of the sky in five bandpasses,
and cataloged a total of over 280 million unique objects, roughly one-third of
which are classified as stars. The vast majority of the photometry is based on
single-epoch observations. Despite this limitation, Ivezi\'c et al. (2005) has shown
that, using color constraints from the high-precision SDSS photometry, an area
in color space can be constructed which allows identification of RR Lyraes with
a completeness of $60\%$ and at an efficiency, defined as percentage of true
variables identified, of 28\%. Even with this rather low efficiency, Ivezic et
al. was able to recover halo substructure features such as the Sagittarius
Stream, and identify other new and potentially interesting overdensities in the
halo.

Multi-epoch observations in SDSS, though covering a much smaller area of sky,
have been shown to be quite efficient for identification of RR Lyrae variables.
Ivezi\'c et al. (2000) used two-epoch SDSS observations by utilizing the overlap area
between SDSS strips. In an area of 97 square degrees they discovered 148
variables with an estimated discovery efficiency of $56\%$. Furthermore,
Sesar et al. (2007) has used multiple observations of SDSS Stripe 82 ($-49^{\circ} <
RA < 49^{\circ}$ and $-1.26^{\circ} < Dec < 1.26^{\circ}$) to identify over 634 
RR Lyrae variables with $g < 20.5$. Using known variables discovered from the
SDSS Light-Motion-Curve Catalog (LMCC) by Bramich et al. (2007), Sesar et al. were able
to determine that the completeness of the multi-epoch Stripe 82 detections was
$\sim 95\%$ with a discovery efficiency of $\sim 70\%$.
The candidate variables in Sesar et al. were initially identified in the LMCC by
comparing their positions to the list of RR Lyrae variables presented by
DeLee et al. (2006). The lightcurves of the remaining candidates were then analyzed
to separate the RR Lyrae variables from other types of variable and non-variable
stars. These were confirmed and Bailey typed using a period-amplitude diagram
similar to what was done in De Lee et al.  These procedures are being followed
to extend the light-curve coverage for these (and fainter) RR Lyraes, based on
the fall 2006 (and eventually) fall 2007 SDSS Supernover Survey campaigns.

In addition to its imaging program, SDSS/SEGUE has now made (as of DR-6;
Adelman-McCarthy et al. 2007), over 218,000 spectroscopic observations of stars
with spectral-type earlier than M. Although clearly a smaller number than that
of the photometry survey, these observations cover the same
footprint as the photometry survey and constitute a second epoch of observation.
It is well known that the hydrogen Balmer lines in RR Lyraes
undergo significant changes in their breadths between minimum
and maximum light (Smith 1995). For RRab variables this change ranges from
spectral type F6 to A8, while it is less pronounced in RRc type variables (F1 to
A8). This
difference in Balmer-line strength compared to $g-r$ color can be used to identify
RR Lyrae variables for those cases where the photometry and spectroscopy
observations are taken out of phase, which will often be the case unless
special steps are taken to coordinate the phases of the observations.

In this paper we describe a technique for identifying RR Lyrae variables from
the combination of spectroscopic Balmer-line strengths and $g-r$ colors, based on
data taken at random phases. The basic technique is discussed in \S 2. In \S 3
we use the RR Lyrae identifications from the LMCC, and additonal photometric
observations obtained from McDonald Observatory, to quantify our
completeness and discovery efficiency. We characterize the entire RR Lyrae
sample in \S 4. Section 5 explores our ability to recover known halo
substructure based on this sample. A brief summary and conclusions are provided
in \S 6.

\section{The Technique}


This technique uses two-epoch observations, one from the single-epoch photometry
and the second from follow-up spectroscopy. Note that many of the spectroscopic
observations were performed on stars that occupy the color region associated
field horizontal-branch stars (e.g., Sirko et al. 2004). The goal is to
identify variable stars that have these two observations taken out of phase with
one another (that is, the suspected variable is observed at different points in
its pulsation cycle). When this occurs, the measured $g-r$ color will be
inconsistent with that that expected from the width of the Balmer lines
determined from the spectroscopy. Obviously, this technique is not able to
recover variables for which the two epochs are taken (by chance) in phase with
one another. 

As is clearly shown in Figure 1.10 of Smith (1995), the hydrogen Balmer lines
in RR Lyrae variables undergo an enormous change in strength between minimum
and maximum light. For the RRab variables the spectral type can change from F6 at
minimum light to A8 at maximum. The variation for RRc variables is more modest,
from F1 to A8. The color of the star, as shown for Johnson
$B-V$ colors in Figure 1.9 of Smith, changes by several tenths of a magnitude
during the full phase of variation. 

In order to explore the relationship between the Balmer-line width and SDSS
$g-r$ color, we have first constructed a grid of synthetic colors and Balmer-
line widths. The colors were determined using Kurcuz Atlas9 (Kurucz 1993) flux
models convolved with the $g$ and $r$ bandpasses from the SDSS (Strauss \& Gunn
2007).
Synthetic
spectra were computed using the Atlas9 model atmospheres and the spectral
synthesis routine SPECTRUM by Gray (Gray \& Corbally 1994). The normalized spectra were
computed at a very high dispersion (0.02 \AA/pixel) and smoothed to the
resolution of the SDSS spectra ($R = 2000$). The Balmer lines H$\delta$,
H$\gamma$, and H$\beta$ were then fit with a Voigt profiles, and both the
equivalent width (EW) and D$_{0.2}$ width (width of the line at $20\%$ below the
local pseudo-continuum) were determined. In order to assemble a more robust
Balmer-line diagnostic, which is particularly needed when using lower
signal-to-noise SDSS spectra, the Balmer-line widths were averaged to obtain
a single EW and D$_{0.2}$ for each T$_{\rm eff}$, log g and [Fe/H]. In averaging
the three Balmer-line widths, the scatter in the final averaged line width was
less than $2\%$ for both EW and D$_{0.2}$ throughout the grid. 
The final grid includes 756 grid points, ranging over 5500~K $ \le  T_{\rm eff}
 \le $ 9750~K; $2.0 \le $ log g $\le 4.5$ ; $-3.0 \le {\rm [Fe/H]} \le 0.0$.

To set model boundaries for the ``non-variable" stars we use the widest possible
range of Balmer lines for a given $g-r$ color. At the lower boundary this was
found to be log g = 2.0 and [Fe/H] = $-3.0$, while the upper boundary is
log g = 4.5 and [Fe/H] = 0.0. This range in log g encompasses virtually all stars on the
horizontal branch, as well as high-surface-gravity main-sequence and blue
straggler stars.

The line widths for stars in DR-6 were computed in the same way as that of the
model data. Figure 1 shows the model D$_{0.2}$ and EW determinations as a
function of $g-r$ color. The bounding lines are shown as the red solid lines;
the lower line corresponds to log g = 2.0, while the upper line corresponds to
log g = 4.5. These lines were generated using grid values, which were then fit
with $7^{th}$ order polynomials. The coefficients for the fits are listed in
Table 1. The black points are 52,315 stars from SDSS DR-6 that fall no more
than $1\sigma$ above or below the bounding lines. These are considered
non-variable stars. It is interesting to note the spread in data points for $g-r
< -0.15$. This is the separation of blue horizontal-branch stars (lower trend)
and main-sequence-gravity blue stars (upper trend).

The green data points with error bars are the stars that are found to be
inconsistent with the non-variable bounding lines.  Lack of consistency 
is determined by demanding that the data points are more than $1\sigma$ outside
of the bounding lines for the D$_{0.2}$, EW, and $g-r$ values. The total number of
green data points is 5,931. (Note that the apparent overlap of green
error bars with the lower bounding line near $g-r$ = 0.25 is only due to vertical
bars at the end of the horizontal error bar.)

In Figure 1, the green data points that appear above the upper boundary line are
stars for which the Balmer-line width is much larger than expected, given the
$g-r$ color. In the case of RR Lyrae variables, stars above the upper boundary
are those for which the spectroscopy was taken near maximum light, while the
$g-r$ was obtained near minimum light. The green data points below the lower
boundary line are stars for which the Balmer-line width is smaller than
expected, given the $g-r$ color, which applies when spectroscopy was taken
minmimum light, while the $g-r$ color was taken near maximum light.

Finally, absent from this plot are a total of 24,704 stars which were culled
from the total sample because they did not pass our minimum criteria for
uncertainty in D$_{0.2}$ and in EW. The criteria for acceptance is a $10\%$
uncertainty in the D$_{0.2}$, and a $15\%$ uncertainty in the EWs, respectively
(hereafter 10/15). Figures 2a and 2b show the uncertainty in the respective line
widths as a function of the D$_{0.2}$ line-width parameter. The acceptance
criteria can be seen as the red dashed line in both plots. These limits where
choosen to include the majority of the locus of points seen near the bottom of
each plot.



\section{The Calibration}

\subsection{Theoretical}

As shown in Figure 1, there are a very large number of stars for which the
Balmer-line widths and $g-r$ colors are found to be inconsistent with one
another. It is crucial to quantify the actual number of stars that we can expect
to truly be RR Lyrae variables from this sample. To begin this procedure we
first look at the theoretical region of parameter space where we would expect to
find the most RR Lyrae variables.

Smith (1995) reports that the mean effective temperature for the hottest RRc variables
is T$_{\rm eff} \sim 7400$~K, while the mean for the coolest RRab variables is
T$_{\rm eff} \sim 6100$~K. We have constructed a theoretical bounding box which
represents the Balmer-line widths and $g-r$ colors for these two limits. The box
was determined using an actual range of 6000~K to 7500~K with parameters drawn from
our grid of theoretical parameters; the box is shown in green in Figure 3. This
particular box uses a log g = 2.5 and [Fe/H] $= -1.5$. We have experimented with
a range of surface gravities and abundances, but these variations lead to only
very modest shifts in the box location.

From this excercise it is clear that a large percentage of the RR Lyrae
candidates from Figure 1 are misidentified, since they clearly lie well
outside of the theoretical confines of the instability strip.  We therefore
have placed more stringent constraints on the identification of the
RR Lyrae sample. In Figure 3 the new sample, shown in black, was selected using
a criteria of $5\%$ uncertainty in D$_{0.2}$, $10\%$ uncertainty in EW, and
uncertainty in $g-r$ less than 0.03 magnitudes (hereafter 5/10). As before, to
be considered a variable candidate, a star must be lie more than $1\sigma$
outside the bounding lines for D$_{0.2}$, EW, and $g-r$, simultaneously. The total
number of stars found using these criteria is 2,142.

Even with our more stringent criteria, it is clear that there remains a
significant number of candidates that we would not interpret to be variable
stars. The obvious region is the large number of stars above the log g = 4.5
bounding line and above the theoretical bounding box. A portion of these stars
are likely to be stars with log g $ > 4.5$. Blueward of $g-r = 0.1$, the Balmer-
line width grows rapidly for main-sequence-gravity stars; stars with log g $>
4.5$ would be expected to lie noticeably above the boundary at this point. The
other, less noticable, region of concern is below the log g = 2.0 boundary, near
$g-r = 0.25$. In this region the Balmer-line diagnostic is losing its
effectiveness. Furthermore, the number of stars in this region skyrockets as we
include more stars located near the main-sequence turnoff. In this region, the
chance of misidentifications is quite high, arising from random scatter among
the tens of thousands of turnoff stars.

\subsection{Empirical Calibration: SDSS Stripe 82}

In a recent publication, Sesar et al. (2007) compared the multi-epoch photometry
observations of Stripe 82 (S82) in the SDSS to 613 variable objects identified
in the SDSS LMCC. Of the RR Lyrae variable stars in this sample, 220 have
follow-up SDSS spectroscopy available, and offer an excellent sample with which
to test our variability identification technique. In total, there are 298
spectra available in DR-6 for the 220 stars. Fifty stars have multiple spectra
(between 2 and 5 spectra); the remaining sample of 170 stars have single
spectroscopic observations.

For this calibration sample we are using the best re-run photometry from the psf
magnitudes, with extinction corrections. The photometric measurements are
identical for all of the spectroscopically observed stars. For a typical
analysis of non-variable stars, we would usually either average the stellar
parameters from mulitple spectroscopic observation or choose the best available
data to analyze. In the case of RR Lyrae variables, the spectral parameters are
changing throughout the phase of pulsation (see \S 4), which removes the option
of averaging the derived parameters. Since the multiple spectra for a given star
were taken at random epochs in the pulsation phase, we choose to begin our
comparsion analysis by treating each spectrum as if it were a unique star (N =
298). We further test our discovery efficiency from the sample of singly
spectroscopically-observed stars (N = 170). As will be shown below, the choice
of the analysis approach has little effect on the determined efficiency.

Figure 4 is a plot of the known RR Lyrae variables, with black dots representing
RRab types and blue dots representing RRc types. The green bounding box is our
theoretically predicted range for the RR Lyrae instability strip. It is
immediately obvious that the theoretical bounds encompass a large percentage of
the known variable stars. We have therefore choosen to limit the parameter space
for RR Lyrae identification in order to minimize the amount of contamination
from misidentified variables.

Given the partial spectroscopic follow-up observation in DR-6, it is difficult
to fully address the final completeness of this sample. Of the 416
identified RR Lyrae variable stars found in the LMCC, only 220 have
spectroscopic observations. Because of the concentrated observational efforts
for objects in S82, this is probably not fully representative of the
spectroscopic completeness of the rest of DR-6. Furthermore, our technique only
identifies those variables that have spectroscopic and photometric observations
that are taken out of phase; spectroscopic and photometric observations which by
chance have been taken in phase will be missed. Finally, because of the
uncertainty criteria placed on the observed data (discussed above) the
completeness of the sample decreases dramatically as one approaches $g= 19$.

Our primary goal is to maximize the discovery efficiency, quantified by the
percentage of true variables to identified candidates. To this end, we
choose to re-scale our theoretically-determined instability strip. The final
limits are set in part by the empirical data of Sesar et al. (2007), and in part by the
desire to avoid large numbers of non-variable stars that would otherwise be
scattered (due to observational error) into the limits of the region we
consider to be occupied by true variables. Figure 4 shows the new limit box as
black dashed lines. The color limit was extended to the blue in order to
encompass several more identified variables. The sloped limit on the cool end
helps to eliminate the enormous numbers of turn-off stars that reside in this
portion of the parameter space. Finally, we did not increase the upper boundary
of the box to include the few RRc variables near $g-r$ = 0.05 because of the
significant number of non-variable stars in this region of the plot that have
Balmer-line widths consistent with log g values greater than our imposed limit
of log g = 4.5.

In the analysis of completeness and efficiency that follows, we consider the
sample of stars in S82 that fall in the color range of $-0.4 < g-r < 0.35$. Our
procedure then selects RR Lyrae candidates, which are then compared to the LMCC
spectroscopic sample. We expect the completeness of the LMCC sample in S82 to be
close to $100\%$. 

Table 2 shows the completeness for various subsamples in S82, using
the 5/10 and 10/15 uncertainty criteria described above. The recovery rate for
the entire sample of S82 variables is quite modest ($31\%$) when the 
10/15 criteria are used. This low completeness is primarily caused by the larger
fraction of stars for which the photometric and spectroscopic observations were
taken in phase. The percentage of stars which cannot be detected using this
technique is $51\%$, which is not surprising. Inspection of Figure
1.11 of Smith (1995) shows that the percentage of time an RRab variable spends
with spectral type F4-F6 (minimum light) is approximately $65\%$. Using this
relation, we would expect to have two unique observations taken at minimum light
a fraction equal to 0.65 x 0.65 = 42\% of the time. The in-phase,
maximum-light probability would be 12\%. Therefore, the percentage of time that
two observations will be taken in phase is expected to be $54\%$ (the
out-of-phase fraction being 46\%), which is consistent with our findings.

It is informative to explore the expected completeness within the confines of
our instability limits, but also considering the out-of-phase detection
threshold. It is clear that the more conservative 5/10 criteria recovers a
significantly lower percentage of RR Lyraes ($\sim42\%$) compared to the 10/15
criteria ($\sim58\%$). This result also applies when we split the sample into
RRab- and RRc-type variables. The majority of the non-recoveries occur very near
the model boundaries, where the $1\sigma$ criteria for detection, in both
D$_{0.2}$ and EW, result in a modest number of losses.

Table 3 shows the percentage of correct identifications for the S82 sample.
Based on our instability-limit criteria, the efficiency of the 5/10 and 10/15
analyses are found to be $\sim54\%$ and $\sim49\%$, respectively. This improves
to $\sim70\%$ for sample stars with $g < 18$. It is also interesting to note
that there were no positive detections for $g < 19$ using either of these
criteria; our selection efficiency for stars fainter than this limit is likely to be 
very low.

To improve our selection efficiency, we next explore the color space of the S82
sample. Figure 5a shows $u-g$ as a function of $g-r$ for the candidate variable
stars in S82 identified with the 10/15 criterion. The small filled circles are
the predicted variables from our procedure, while the large circles represent
the known RR Lyrae variables that fall in the identification areas. It is
immediately clear from this plot that a significant number of misidentified
candidates inhabit the region of the plot where $u-g < 1.0$. The boxed region in
the plot is the area defined by Ivez\'ic et al. (2005), which has been shown to
encompass a color space that represents nearly $100\%$ completeness for RR Lyrae
variables. Although Figure 5a shows that nearly all variables fall within these
bounds, there remains a small percentage of misidentified variables in the range
$0.98 \le u-g \le 1.03$. Since we are primarily concerned with detection
efficiency, we use the color cut $u-g \ge 1.03$ to identify candidate variables.
Although this choice may remove a small percentage of RR Lyrae from our sample,
it will more importantly remove a significant number of misidentifications.

Figure 5b shows further reasons for our choice of color cuts.  The large red circles
in this Figure are the entire spectroscopic sample of known variables in S82.
The small black circles are our entire DR-6 sample of candidates identified by
the 10/15 criterion. It is clear from inspection that very few RR Lyrae
variables inhabit the bottom portion of the Ivez\'ic et al. bounding region. The
full candidate sample, however, has a similar density of stars in this region as
in the area immediately above. This suggests that the vast majority of the
stars below the $u-g = 1.03$ cutoff are in fact misidentifications. By
application of this color cut, Table 3 shows that our identification efficiency
jumps to $84\%$ and $77\%$ for the two criteria. For stars with $g < 18$,
the detection efficiency improves to $91\%$ for the $5/10$ criterion.  Although
this efficiency is slightly higher than found for the 10/15 criterion, the
completeness for the 10/15 sample is markedly higher than for the $5/10$
criterion.  Thus, we choose to use the $10/15$ criterion for the remainder
of this paper.  Figure 6 is a final plot of the S82 sample, constructed from
application of the 10/15 criterion along with the $u-g$ color cut. The green
error bars represent the candidate variables for S82. Very few
misidentifications remain in the sample shown.

We now test the efficiency of variable detection by comparing the candidate list
from S82 to the smaller subsample of stars which have only one spectroscopic
observation (N = 170). The results based on application of the two criteria
(with color cuts) are listed in Table 4. Although the efficiency is now slightly
lower, the two samples still have around $75\% - 80\%$ efficiency for stars with
$g < 19$. This value should be considered a lower limit on the procedure
efficiency, because we have used the original S82 sample for our analysis, but
have removed correctly-identified stars with multiple spectroscopic
observations.

As a final test of our procedure, we have allowed the inclusion of stars with
multiple spectroscopic observations, but have only counted the positive
identifications {\it once} for a given star. The results of this test are listed
at the bottom of Table 4. We have again recovered an efficiency similar to our
first attempt, in excess of $80\%$ for the sample with limiting
magnitude of $g = 19$. Furthermore, all tests suggest that a sample of variable
stars with $g < 18$ and selected with the 5/10 criteria will suffer no more than
$10\%$ to $15\%$ misidentification.

\subsection{The McDonald Observatory Sample}

To further test our procedure we have initiated a program to identify variable stars
selected from SDSS DR-5 using our routines. The ultimate goal of this program is
to construct full light curves for stars of particular interest in the study of
halo substructure. We present here our first efforts at identifying true variable
stars from the candidates we have selected. This sample is of interest because
it is not confined to Stripe 82, but rather, drawn from the full SDSS DR-5 database.

Data from three separate runs have been obtained at McDonald
Observatory using the 0.8m telescope with the Prime Focus Corrector (PFC). The
PFC uses a Loral Fairchild 2048 by 2048 pixel CCD, with a pixel size of
$1.35''$, and a resulting field of view of $46.2'$ by $46.2'$. Smaller
sections of the chip were used to reduce the readout time on observing runs
dedicated to observing RR Lyrae candidate stars. All candidates were observed at
airmass less than 1.6, and with integration times ranging from 100 to 300 seconds.
Whenever possible, multi-filter observations (either Johnson BV or VR) were made
in order to test variability in both bandpasses.

All of the data were processed using standard reduction techniques,
including bias and flat field corrections. The bulk of the reductions were
performed using CCDPROC in IRAF\footnote{IRAF is distributed by the National Optical 
Astronomy Observatories,
which are operated by the Association of Universities for Research
in Astronomy, Inc., under cooperative agreement with the National
Science Foundation}. The RR Lyrae
candidate stars observed were sufficiently isolated to allow aperture photometry
to be performed using the commercial package MIRA \citep{nee92}. Differential
photometry was then performed, using the resulting instrumental magnitudes. A
selection of stars was taken from each field to find stars to use as
comparisons. Standard deviations were calculated from the magnitudes of these
stars, and those with the least scatter were selected to use as comparison
stars. The average magnitude was calculated for each star from their measured
magnitude on each image, and the offset from this mean was found for each
standard. The offsets for all standards on a given image were then averaged to
obtain a correction for that image. The offsets were then applied to the
measured magnitudes for the suspected variable and selected comparison stars for
each image. The results were plotted in order to verify that the comparison
stars exhibited non-variable light curves. If the light curves suggested that
the values for a given image systematically varied from a flat curve, a second
average was calculated, and the offsets were found and applied again. Two
iterations proved sufficient in order to flatten the curves of the comparison
stars.

We were not concerned with obtaining calibrated magnitudes, since
any stars shown to be variables are being targeted for follow-up observations to
obtain complete light curves. The procedures described above removed the effects
of differing air mass or changing observing conditions, and allowed us to achieve our
present goal of the verification of variability.

While the light curves alone could be used to make some statement of
variability, we wanted further confirmation. We adopted the statistical analyses
detailed in Sesar et al. (2007) to make more quantitative measures
of the variability of the candidate stars.  Sesar et al. searched for RR Lyraes in
Stripe 82 of the SDSS. Each star they observed generally had a small number of
observations, taken at random times (driven by the SDSS Supernova Survey
cadence), so rather than attempting light curve fitting they employed low-order statistics
to determine variability. Our data is not obtained in the same manner, but, like
theirs, it does not provide sufficient observations to obtain full light curves,
so we chose to use their statistical approach.

Sesar et al. used two criteria to choose variable stars from their large data
set, which appear as equations 1-4 in their paper. Their criteria for selection is:

${\sigma(\nu)\ge 0.05\; {\rm mag}}$

${\chi^2(\nu)\ge3},$

\noindent where $\sigma$ is computed from the RMS deviation of the individual measures
and the average photometric errors, ${\xi(m)}$, and ${\chi^{2}}$ is computed from
the variance in each observation divided by the individual photometric
uncertainties; this quantity is used to test the departure of the measurements
from a Gaussian distribution of errors.

Terms were calculated for each color, where available, to further confirm
variability. For our calculations, we used the empirical photometric error
resulting from our aperture photometry as $\xi$. Further errors introduced from
atmospheric changes were neglected, since these were suppressed if not
eliminated by the process of flattening the curves of the comparison stars.

Among the limited observations we were able to carry out in this initial
investigation, there were 15 stars that we could confidently assign into
variable or non-variable classifications. Four additional stars require
additional data in order to confirm their status. For these stars, the formal
statistics indicate they are not variable, but the partial light curve suggest
otherwise. These stars will be observed again on a later date, so that
we can make a final decision on their variability status. The results are listed in Tables
5, 6, and 7. Figure 7 shows a characteristic light curve for one of our variable
stars, and one lightcurve for a non-variable object.

Of the 15 stars that we could confidently classify, 7 are variable. Of
these seven, 2 fail to exceed the $\chi^2 > 3$ criterion in one or both colors.
Star SDSS J165340.87+342302.8 has a $\chi^{2}$ in
V of 2.641, and a $\chi^{2}$ in B of 6.064. The partial light curve exhibits
clear changes, so we designate it as variable. SDSS J130707.52+580039.2 
has
$\chi^{2} = 2.821$ in V and 2.845 in R. Again, when considering the partial
light curve, we confidently identify the star as variable. This may indicate
that $\chi^{2} > 3$ is an overly conservative criterion.

Figure 8 shows a plot of D$_{0.2}$ vs. $g-r$ for the 7 confirmed candidates and the 8
stars that show no signs of variability. All 7 of the variable stars fall within
the region of this plot for an expected variable detection. Of the 8
non-variable stars, 3 fall within the bounding lines for non-variability. These
stars were mistakenly chosen to be candidates because of an error in the Balmer-
line averaging software in an early version for the DR-2 sample. Although they
are mistakes, we have included them in the sample because they are found to
indeed be non-variable. One non-variable star lies above the limiting bounds of
our instability bounding box, which we did not implement prior to observing.
Four of the non-variables were misidentifications at the telescope. It is
interesting to note that 3 of these stars lie in the region where we might
expect to find more RRc variables with low amplitude oscillations that may have
escaped our small number of observations ($N = 5-7$). The final result for the
McDonald study is 7 out of 11 variable detections, for an efficiency of $63.6\%$.

\section{Analysis of Sample}
\subsection{The Multiple Spectra Sample}

As discussed by Smith (1995), the Balmer-line strength increases as an RR Lyrae
variable star approaches maximum light.  Also, as is shown in Smith's Figure
1.12, the radial velocity, due to the expansion of the photosphere near maximum
phase, quickly changes from positive values to negative values ($\sim -40$ km
s$^{-1}$) in RRab variables. It is therefore expected that a variable star's
radial velocity and Balmer-line widths should exhibit an anti-correlation during
the pulsation phase.

To explore this effect in the DR-6 sample, we have chosen a set of 19 stars that
have more than three spectroscopic observations. According to the
LMCC classifications, this sample has a total of 13 RRab and 6 RRc variables.
Figure 9a shows the radial velocity for the RRab sample as a function of the
D$_{0.2}$ line width. The solid line is a linear regression to the data points
for each star, and exhibits a general decrease in radial velocity as the line width
grows. A similar effect is seen for the RRc variables shown in Figure 9b, but to
a much lesser extent, with more modest changes in both line width and radial
velocity.

We use linear regressions to map the overall changes in line width and
radial velocity for both the RRab and RRc variables. Figure 10 shows the trends
for both types of stars. Although any single star with data taken at a random
phase will not exhibit the full extent of the expected changes, the entire group of
RRab exhibits an overall change in line width of $\sim 12 \AA$, and an overall
change in velocity of $\sim 75$ km s$^{-1}$, in keeping with expectations. The RRc
sample exhibits a much smaller range in Balmer-line widths, and a nearly
constant change in radial velocity of $\sim -20$ km s$^{-1}$. This is a smaller
range in radial-velocity variations that might be expected for RRc variables,
however, one variable does exhibit an overall change of $\sim -45$ km s$^{-1}$, in
keeping with that expected for RRc variables from minimum to maximum light
(Smith 1995).

These results strongly suggest that the DR-6 data is of sufficient quality to
detect overall differences in the radial velocity data due to pulsation, and that
the sense of the shift will be anti-correlated with the change in Balmer-line
width.

Our procedure samples two regions in the Balmer-line width, $g-r$ color
plane. The upper region is consistent with the Balmer lines being observed
near maximum light and the $g-r$ color index being measured near minimum light.
The lower region is the reverse. It is therefore expected that the two regions
should show an overall average difference in radial velocity, with the upper
region having a more negative velocity than the lower region. We test this by
again returning to the LMCC sample, and examining multiple spectra of stars that
are detected in either the upper or lower region, and also in the bounding
non-variable region. This sample contained no stars with spectral observations
in both the upper and lower detection areas. We therefore chose to compare the
in-phase velocities to the velocities from the out-of-phase detection areas. The
average difference in velocity for the upper region minus the in-phase spectra
($N = 8$) was $<v> = -29.7$ km s$^{-1}$, while the lower region minus the
in-phase spectra ($N = 9$) was $<v> = +27.9$ km s$^{-1}$.  

The above result suggests that we should find an average difference between the upper
and lower detection areas when comparing our entire sample of stars. The expected
amplitude of this average difference is difficult to predict because the in-phase stars from 
the above analysis can be at any random pulsation phase.  The above exercise  
suggests that we can expect average difference on the order of 30 to 50 km s$^{-1}$. 

In order to test for radial-velocity variability in our full sample, we explore the
average difference in velocity splits between the upper-region and lower-region,
out-of-phase detection areas. Figure 11 shows the cumulative normalized
distribution of the velocities with respect to the Galactic rest frame, v$_{\rm
gsr}$, for the two samples. This Figure clearly indicates a tendency for the
stars chosen from the upper-region candidate sample to be shifted to more negative
velocities. We interpret this shift as due to the negative velocity of pulsation
experienced near Balmer-line maximum, which is superposed on the star's systemic
velocity.

For the upper-region sample ($N = 391$), $<v_{gsr}> = -26.9$ km s$^{-1}$, with a
standard error of the mean $\pm 5.6$ km s$^{-1}$, while the lower-region
sample has an average velocity of $<v_{gsr}> = 1.4$ km s$^{-1}$, with a standard
deviation of the mean of $\pm 4.3$ km s$^{-1}$. There thus exists a nearly
$3\sigma$ difference between the upper-region and lower-region samples, in the
sense expected. Furthermore, the amplitude of the effect is 28.3 km s$^{-1}$,
which is indistinguishable from the expected lower limit set by the multi-epoch
spectroscopic sample. Finally, the velocity dispersion of both samples are
virtually identical ($\sigma = 112$ km s$^{-1}$), suggesting that both are drawn
from the same parent population of stars.

As a final test of variability in the full sample we search for changes in
Balmer-line widths and velocities for all of the DR-6 stars that have multiple
spectroscopic observations. In the S82 sample there are a total of 15
known RR Lyrae variables that have multiple spectral observations, and
have been classified by our technique to be variable stars. In this sample, 9
stars exhibit greater than $1\sigma$ variations in EW, D$_{0.2}$, and velocity
($1\sigma = 10$ km s$^{-1}$). If we use this criteria as a spectroscopic
detection of variability, $60\%$ of the 15 stars chosen from the S82 sample
would be classified as variables.

We applied the above spectroscopic variability criteria to the
DR-6 sample of 39 stars with multiple spectroscopic observations that are
located outside of the S82 footprint. Of these, 23 ($59\%$) show variability in
the spectroscopic features, similar to the percentage found for the known RR
Lyrae sample in the S82 region. We also note that 11 of the 23 stars in the full
DR-6 sample showed departures greater than $3\sigma$, making them prime
variable-star candidates.  Given these results, and the aforementioned
velocity shifts between the upper- and lower-region samples, we expect that this
efficiency of discovery for RR Lyrae variables is maintained throughout the
full DR-6 candidate sample.

Table 8 lists the entire sample of RR Lyrae candidates chosen with the 10/15
selection criteria.
For stars that have multiple spectroscopic observations we
have averaged the final heliocentric radial velocity. These stars are indicated by
the number of spectroscopic observations. This Table also lists the stars which were
identified using the more restrictive 5/10 selection criteria, for readers
interested in conducting follow-up light curve observations. As noted in the
calibration section above, for stars with $g < 18$ we expect $90\%$ discovery
efficiency for the 5/10 sample.

\subsection{Distances}

One strength of our RR Lyrae sample, selected as above, is the high percentage
of correctly identified variables located at large distances. Our sample has
excellent potential to identify halo substructure on the basis of distance and
kinematics, and to probe kinematics along stellar streams.  

RRab variables can change their apparent visual magnitudes during their
pulsation cycles by as much as 1.3 magnitudes (in $V$), while the change in RRc
variables is a more modest 0.5 magnitudes. Because of these large variations,
distances to RR Lyraes are typically computed using the average magnitude of the
star, based on either a simple average of the maximum and minimum magnitudes, or
an arithmatic mean over the entire pulsation cycle. Use of a single magnitude
measurement, taken at a random phase, can introduce significant error in the
derived distance. Considering the full variation in apparent magnitude from peak
to peak, the effect on the derived distance could be as large as $\sim45\%$ for
stars with the typical brightness of our sample.  Since our sample
has only one photometric observation, it is not possible
to directly compute an average apparent magnitude for an individual star.
Instead, we have developed a simple magnitude estimation procedure, based on a
star's measured $g-r$ color relative to the $<g-r>$ for the entire sample.

Our estimation procedure makes use of the fact that a variable's $g-r$ color
changes as a function of the pulsation phase, with the bluest color occuring near
maximum light and the reddest near minimum light.  We adopted the simple
assumption that a one-to-one correspondence exists, for a given variable,
between its $<g>$ magnitude (averaged over its pulsation cycle) and its $<g-r>$
color. Although the color limits for a given variable depends on its location
within the instability strip, and in part on its metallicity, we make a further
simplifing assumption that all variables in our sample cover the same range in
$g-r$, and that this range is set by the empirical limits imposed by our
selection criteria. These two assumptions essentially allow us to treat our
sample as a single variable observed at various points during its pulsation
phase. We then use the $<g-r>$ for the entire sample to derive the observed
color shift of a given variable from the sample mean. This shift is used to
calculate an expected offset from the average $<g>$ magnitude of the star, which
can be applied to the observed $g$ magnitude to obtain an estimate of the true
$<g>$ of the star.

In practice, we have also separated potential RRab and RRc
variables in our sample. RRc variables are systematically bluer than RRab
variables, so the use of $<g-r>$ for the complete sample would introduce a
systematic offset. Inspection of the LMCC variables in Figure 4 shows that the
RRc variables are primarily confined to the upper left-hand corner of the
instability region, so we use this region to construct limits for the RRc stars
in our sample. The D$_{0.2}$ and $<g-r>$ limits of the RRc bounding box are
given by:

{$14.00 <$ D$_{0.2}$ $< 21.77$}

{$-0.17 < g-r < 0.14.$}

We next test our procedure to see if the simplifying assumptions we have made
predict average magnitudes that more closely represent the $<g>$ magnitude. 
We use the simultaneous Cousins $BV$ observations of four well-known variables
(BH Peg, SS Leo, UU Vir, and X Ari) reported by Carrillo et al. (1995). These RRab
variables were observed more than 180 times, and span a wide range in metal
abundance ($-2.4 < {\rm[Fe/H]} < -0.8$). Figure 12 shows the Carrillo et al.
observations. It is immediately clear from inspection of this Figure that a
strong correlation exists between the $V$ magnitude and $B-V$ color for all four
stars. The slope of the linear fit to each star exhibits a very small range (2.1
to 2.4), which indicates that that the $<B-V>$ color can be used to predict the
$<V>$ magnitude. It is also clear that color shifts exist between stars within the
sample.  The stars X Ari and BH Peg inhabit a redder color range
than SS Leo and UU Vir. There is also a small hysteresis at the red end of the 
variation between the color of the rising an descending phase.  
When using the average color of the entire sample to
predict an average magnitude for an individual star, such offsets will introduce
systematic errors in the magnitude correction, depending on the location within
the instability strip.

We have tested our procedure using all 813 data points for these four stars. The
$<B-V>$ color was found to be $-0.468$. We also  the arithmatic mean
magnitude of each star individually, and compared the predicted $<V>$ to the
true mean for each star. Figure 13a is a histogram of the difference
($<V>_{predicted} - <V>_{observed}$). The standard deviation for this
distribution is $\pm 0.181$ magnitudes. In comparison, Figure 13b is a histogram
of the difference between the observed $V$ and $<V>$, which is the expected
distribution if no correction is applied to the data. The standard deviation for
this distribution is $\pm 0.635$. The systematic offsets in predicted $<V>$ for
the individual stars range from $-0.13$ to +0.11. Thus, we conclude that
considerable improvement to our sample distance estimates can be achieved, even
if we ignore the color ranges of individual stars. The final computed
uncertainty in distance using the predicted $<V>$ magnitudes is $\pm8.3\%$.

The other source of uncertainty in distance arises from the dependence of
absolute magnitude on metal abundance. One way to correct for this effect is to
determine the abundance using the inverse relationship between the CaII~K and
Balmer-line strengths, as employed by Layden (1994). Unfortunately, we are unable to
determine reliable abundances for our sample because of the need to avoid the
rising phase of pulsation when using this approach (Freeman \& Rodgers 1975). We
therefore estimate the additional uncertainty when this metallicity 
abundance dependency is ignored.

We first adopt the Demarque et al. (2000) absolute magnitude of $M_{V} = 0.55$ at a [Fe/H]
$=-1.60.$ This value is then transformed to an absolute $g$ magnitude, $M_g
= 0.594$, using the transformation equations by Fukugita et al. (1996).  
Clementini et al. (2003)
reports a slope in $M_{V}$ as a function of [Fe/H] of 0.214. Assuming this same
slope applies to the $g$ absolute magnitude relationship, we obtain:

$M_g = 0.214[Fe/H] + 0.94$.  

\noindent If we consider that the range of abundance for a typical halo sample covers 
$-3.0 < [Fe/H] < -0.5$, the additional uncertainty in $M_g$ is no more than
$\pm0.25$. Combining this result with that of the predicted $<g>$,
we estimate that our final uncertainty in distance for typical stars in the DR-6 
sample to be $\pm14\%$.

Finally, as shown earlier in this section, the radial velocity is affected by
the pulsation phase, with the spectroscopic-maximum sample exhibiting a $-28$ km
s$^{-1}$ offset from the spectroscopic-minimum sample. The radial velocity for RR
Lyrae variables is typically assigned at minimum light, therefore we have
applied the above offset to the subsample with spectroscopy taken near maximum
light with the assumption that most of this subsample is also at the 
radial velocity minimum.

\section{Results}

\subsection{Sample Properties}

Figure 14a is a histogram of the predicted $<g>$-magnitudes for the DR-6 sample.
The peak of the distribution occurs at $<g> \sim 17$, and then rapidly declines
for fainter magnitudes. This decline is a result of the strict uncertainty
criteria placed on the sample, in particular on the Balmer-line widths.
Uncertainty in the line widths increases at fainter magnitudes due to the
decrease in signal-to-noise of the spectroscopy. As a result, our sample does
not probe to the faintest available magnitudes in the DR-6 spectroscopic sample.

Figure 14b is a normalized cumulative histogram of the estimated distances for
the sample. The position of $<g> = 18$ and $<g> = 19$ are marked on the
distribution. As discussed in \S 3, the expected efficiency for the 10/15
criteria is $\sim84\%$ for $<g> < 18$, and $\sim82\%$ for $<g> < 19$. The histogram
shows that $81\%$ of the current sample exhibits $<g> < 18$, while $94\%$ of
the sample satisfies $<g> < 19$.  Thus, we expect that nearly the
entire sample has a misidentification rate smaller than $20\%$.  Although the
total DR-6 sample of candidate variables contains only 1087 stars, the relative
purity of the sample allows us to probe kinematic properties that would
otherwise be lost in a larger, but less clean, sample. 

\subsection{Spatial Distribution and the Sagittarius Stream}

Figure 15a is a plot of the Galactic YZ distribution for our sample stars, where
(0,0) is the location of the Galactic center. Inspection of this Figure indicates
that the sample is not uniformly distributed within the DR-6 footprint. Rather,
there are holes and clumps of stars, particularly at distances greater than 10
kpc. These features are over-emphasized in this diagram because of the attention
they received during spectroscopic follow-up. The two most prominent clumps are
labeled as the Southern and Northern Arms of the Sagittarius Stream (larger open
circles). The Southern Arm overdensity is the well-known, dynamically cold,
trailing arm of the stream, while the Northern Arm is near the apogalacticon
point of the Sagittarius orbit. We report on kinematic evidence for both the
leading and trailing Northern Arm in Wilhelm et al. (2007). Figure 15b exhibits similar
inhomogeneities in the Galactic XZ plane. We have again highlighted the
Sagittarius Stream components.

We further explore our ability to recover halo substructure within our sample by
comparing to published analysis of Blue Horizontal-Branch (BHB) stars in the
Southern Stream. Figure 16 is a plot of the heliocentric velocity as a function
of distance for stars that fall within the confines of the Southern Stream
($0^{h} < \alpha < 3^{h}$). It is clear that the sample recovers the cold
trailing arm of the Sagittarius Stream. We measure the mean velocity of this
clump to be $<v> = -157.9$ km s$^{-1}$, with a dispersion of $\sigma = \pm24.1$ km
s$^{-1}$. The mean distance (from the Galactic center) is found to be $<d> =$
27.1 kpc. The distribution of stars in this plot is qualitatively similar in
both velocity and distance to the clump of BHB stars shown in Figure 14 of
Sirko et al. (2004) for the trailing arm. Yanny et al. (2004) also uses BHB stars to trace the
trailing arm, and finds $<v> = -160$ km s$^{-1}$ and $\sigma =
33$ km s$^{-1}$, and an average distance $<d> = $28 kpc. These values are
indistinguishable from those determined using our sample.

Based on this comparison, we gain confidence not only in the identification of the
trailing arm itself, but also with our ability to compute accurate distances and
velocities for the full DR-6 RR Lyrae sample.

\subsection{The Northern Polar Cap} 

In a recent paper, Mart\'{\i}nez-Delgado et al. (2007) predict that the leading tail of
the Sagittarius Stream in the Northern hemisphere is expected to trail down
nearly perpendicular to the Galactic plane. They further argue that the
discoveries of the Virgo Over Density (VOD) by Juri\'c et al. (2005) and the clump of
RR Lyrae stars (VSS) reported by Duffau et al. (2006), with $v_{gsr} = 83$ km s$^{-1}$ and
heliocentric distance of $\sim 19$ kpc, are consistent with the leading and
trailing arms of the Sagittarius Stream, respectively (if an oblate ($q < 1$)
dark matter halo is adopted). Mart\'{\i}nez-Delgado et al. further predict that the
VOD will have large negative velocities if it is actually part of the leading
Sagittarius arm in the Solar neighborhood.

We now use our RR Lyrae sample to examine the Northern Polar Cap (NPC) region
(defined by $b > 70\degr$). Figure 17 shows histograms of the NPC split on
distances from the Galactic plane between $5 < Z < 27$ kpc and $27 < Z < 45$
kpc. It is clear from inspection of this Figure that the distribution of stars
in both samples is not consistent with expectations for the general halo field
population (which has $v_{gsr} \sim 0$ km s$^{-1}$), and velocity dispersion on
the order of 100 km s$^{-1}$. Furthermore, there are only two stars with 
$z < 8$ kpc, so the thick disk contribution should be negligible.  
The more distant Z sample has an average $v_{gsr}
= -48.2$ km s$^{-1}$ with a dispersion of 63.2 km s$^{-1}$. The prominent
negative velocity peak has $<v_{gsr}> = -76.4$ km s$^{-1}$ and dispersion 29.8
km s$^{-1}$, values that are consistent with the expected infalling leading arm
of the Sagittarius Stream.

The more nearby (low Z) sample appears to be a mixture of substructure. Although
there is likely a general field population present in this sample, there are
also recognizable peaks within the velocity distribution. One such peak appears
near $v_{gsr} \sim 100$ km s$^{-1}$ and appears to be consistent with the
velocity found by Duffau et al. for the VSS. There is also a larger peak near
$-100$ km s$^{-1}$ which may be affiliated with the more distant detection of
the leading arm. Lastly, there is a small group of stars with $Z \sim 10$ kpc
which has an even larger negative velocity ($v_{gsr} \sim -200$ km s$^{-1}$), and
may associated with the VOD.

It is important to recognize the similarity of our RR Lyrae data to the models of
Mart\'{\i}nez-Delgado et al. (2007).  The models presented in that paper indicate that the
trailing arm and leading arm will overlap at approximately 20 kpc, and that the
detection of the trailing arm may be the explanation for the positive velocities
found for VSS by Duffau et al. (2006).  From inspection of our data
beyond $Z = 27$ kpc, there are virtually no positive-velocity stars.
Furthermore, the secondary peak found at $\sim 100$ km s$^{-1}$ is only occupied
by stars from the Z $< 27$ kpc subsample. This result suggests that the VSS may
indeed be associated with the Sagittarius tidal stream, as predicted in the
models of Mart\'{\i}nez-Delgado et al. (2007).

Wilhelm et al. (2007) further examines the Northern Arm of the Sagittarius
Stream by identifying both the leading and trailing arms across the range
$110\degr < \alpha < 220\degr$. The results indicate that the negative-velocity
sample is spread over an extended region of the sky, and is consistent with the
expected distance and kinematics of the Sagittarius Stream.

\section{Conclusions}

The technique of using out-phase photometry and spectroscopy to
detect variability was originally developed to identify a small number of
potential RR Lyrae variables stars in the BHB sample of Wilhelm et al. (1999). In the
past, such a procedure was not as practical for identifying large numbers of 
RR Lyrae variable candidates because of the large amount of telescope time
required to obtain spectroscopy. Today, the unprecedented spectroscopic sample of
the SDSS DR-6 has made it feasible to use this procedure to identify a large and
distant sample of RR Lyrae candidates. 

Based on the tests carried out in this paper, it appears that a relatively clean
sample of RR Lyraes (discovery efficiency $> \sim 85\%$) can be achieved using
one photometric and one spectroscopic observation from the SDSS. This is primarily
due to the enormous changes in Balmer-line strength which occur during the
pulsation phase of such stars.

Although the photometry for the candidate variables is taken at a random phase,
we have also shown that it is possible to reasonably predict the average $g$ magnitude
for stars in the sample, in order to limit the uncertainties in derived
distances.  The resulting distances, velocities, and relative purity of the
sample make it feasible to clearly recognize halo substructure, and to
investigate properties of the Sagittarius Stream. Our sample recovers
both the Southern and Northern arms of the Sagittarius Stream, and iallows us to recognize
the signature of the descending, leading tail of the Northern Arm.

Finally, we mention that this procedure can be also be used to recognize RR Lyrae
variables for projects where precise luminosities are paramount. One example is
the use of stellar probes to bracket the distances to high velocity clouds in
the halo of the Galaxy (e.g., Wakker et al. 2007). Such a sample requires
spectroscopy of the stellar probe in order to determine radial velocity, metal
abundance, and distance to the star. Recognition of RR Lyrae variability using
our technique can help constrain the distance to the stellar probes, and
ultimately the distance to the high velocity cloud.

\acknowledgments
Funding for the SDSS and SDSS-II has been provided by the Alfred P. Sloan 
Foundation, the Participating Institutions, the National Science Foundation, 
the U.S. Department of Energy, the National Aeronautics and Space 
Administration, the Japanese Monbukagakusho, the Max Planck Society, and the 
Higher Education Funding Council for England. The SDSS Web Site is 
http://www.sdss.org/.

    The SDSS is managed by the Astrophysical Research Consortium for the 
Participating Institutions. The Participating Institutions are the American 
Museum of Natural History, Astrophysical Institute Potsdam, 
University of Basel, University of Cambridge, Case Western Reserve University, 
University of Chicago, Drexel University, Fermilab, the Institute for Advanced 
Study, the Japan Participation Group, Johns Hopkins University, the Joint 
Institute for Nuclear Astrophysics, the Kavli Institute for Particle 
Astrophysics and Cosmology, the Korean Scientist Group, the Chinese Academy of 
Sciences (LAMOST), Los Alamos National Laboratory, the Max-Planck-Institute 
for Astronomy (MPIA), the Max-Planck-Institute for Astrophysics (MPA), New 
Mexico State University, Ohio State University, University of Pittsburgh, 
University of Portsmouth, Princeton University, the United States Naval 
Observatory, and the University of Washington.

This research was supported in part by NASA, through the American Astronomical
Society's Small Research Grant Program award to RW. RW, TCB, YSL, and ND
acknowledge partial funding of this work from grant AST 07-07776, awarded by the
US National Science Foundation. TCB, YSL, and ND also acknowledge support from
grant PHY 02-16783; Physics Frontier Center/Joint Institute for Nuclear
Astrophysics (JINA), awarded by the U.S. National Science Foundation.

CAP is grateful for NASA (NAG5-13057, NAG5-13147) support.

This paper includes observations obtained at the McDonald Observatory of The University of
Texas at Austin.

\clearpage



\begin{figure}
\epsscale{1.0}
\plottwo{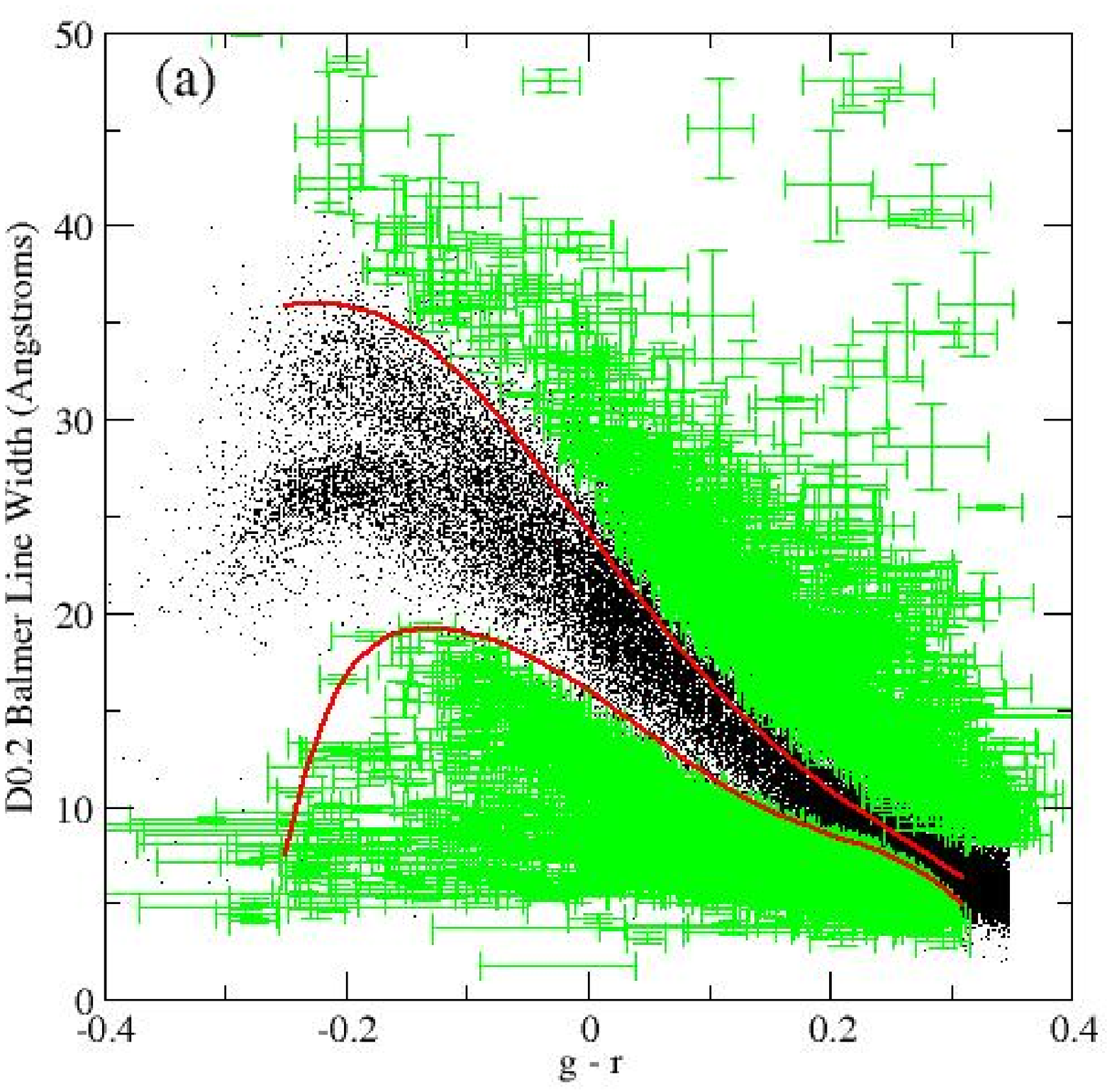}{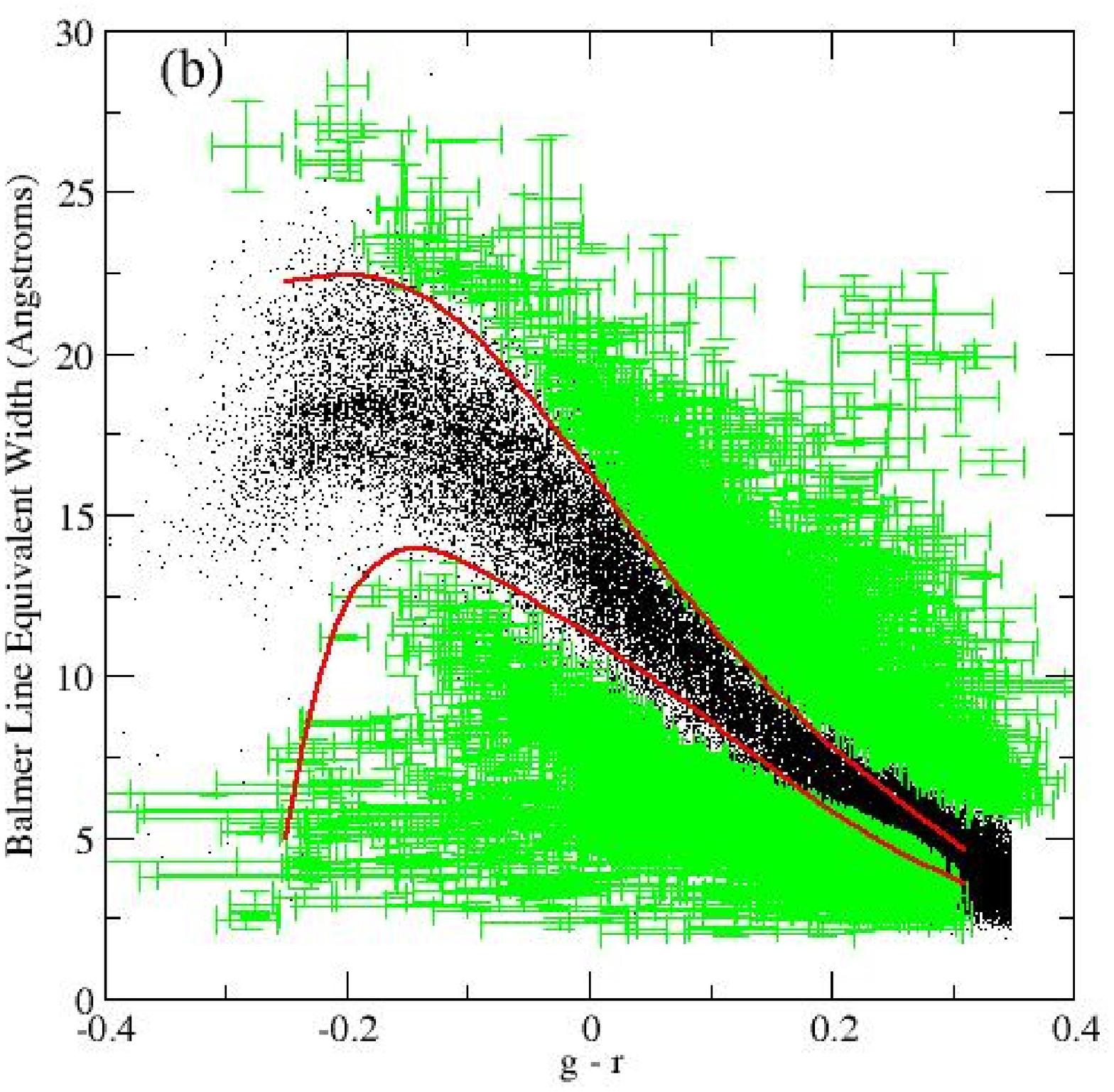}
\caption{SDSS DR-6 sample showing the Balmer-line D$_{0.2}$ width as function of
$g-r$ color (a)
and equivalent width (b). The black dots represent stars which show no variability  while
the green error bars represent the stars that have inconsistency between the 
g - r color and the Balmer-line width. Solid red
lines are the boundaries for normal stars with log g = 2.0 (bottom line) and log
g = 4.5 (top line).
\label{fig1}}
\end{figure}


\begin{figure}
\epsscale{1.0}
\plottwo{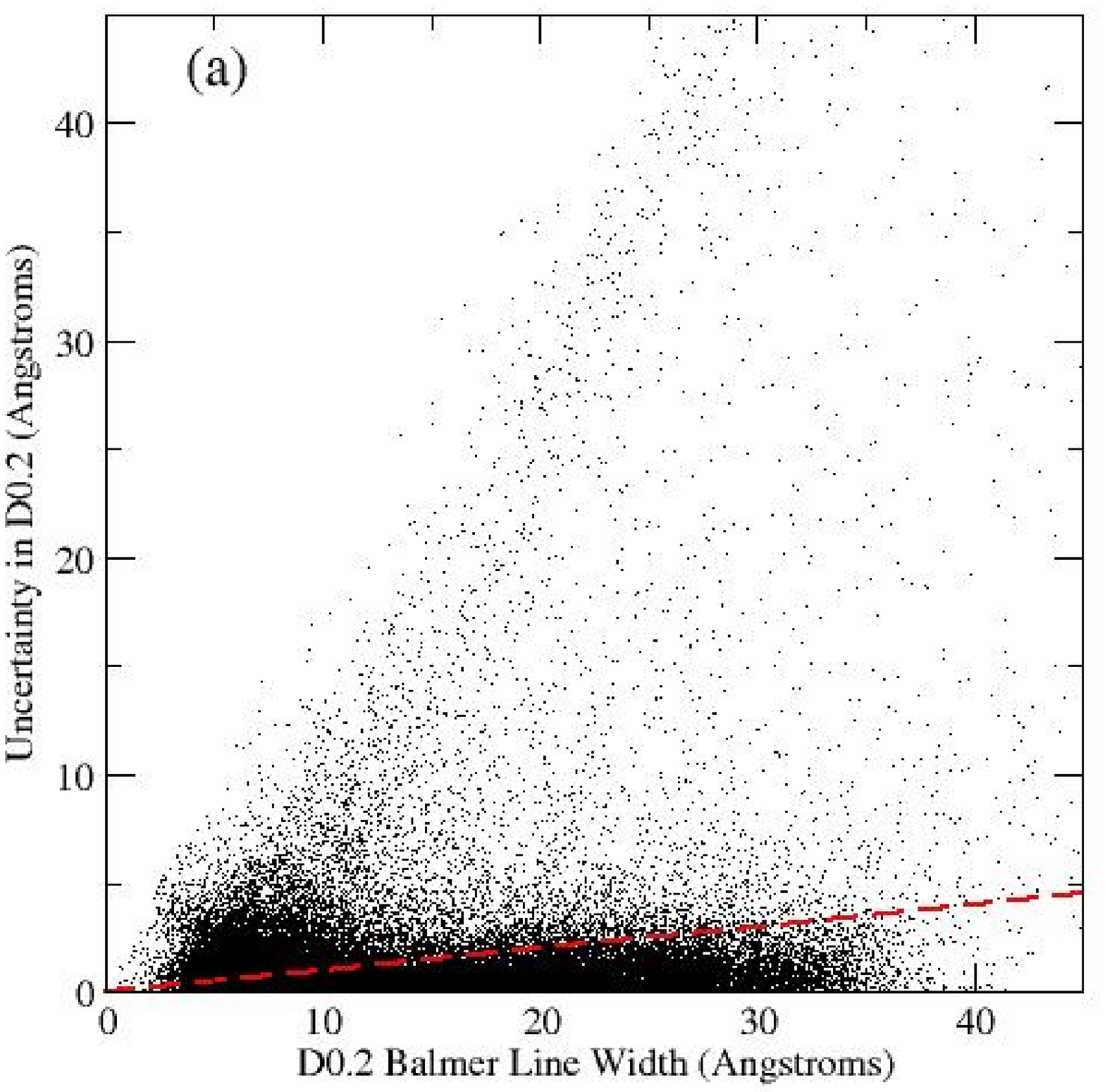}{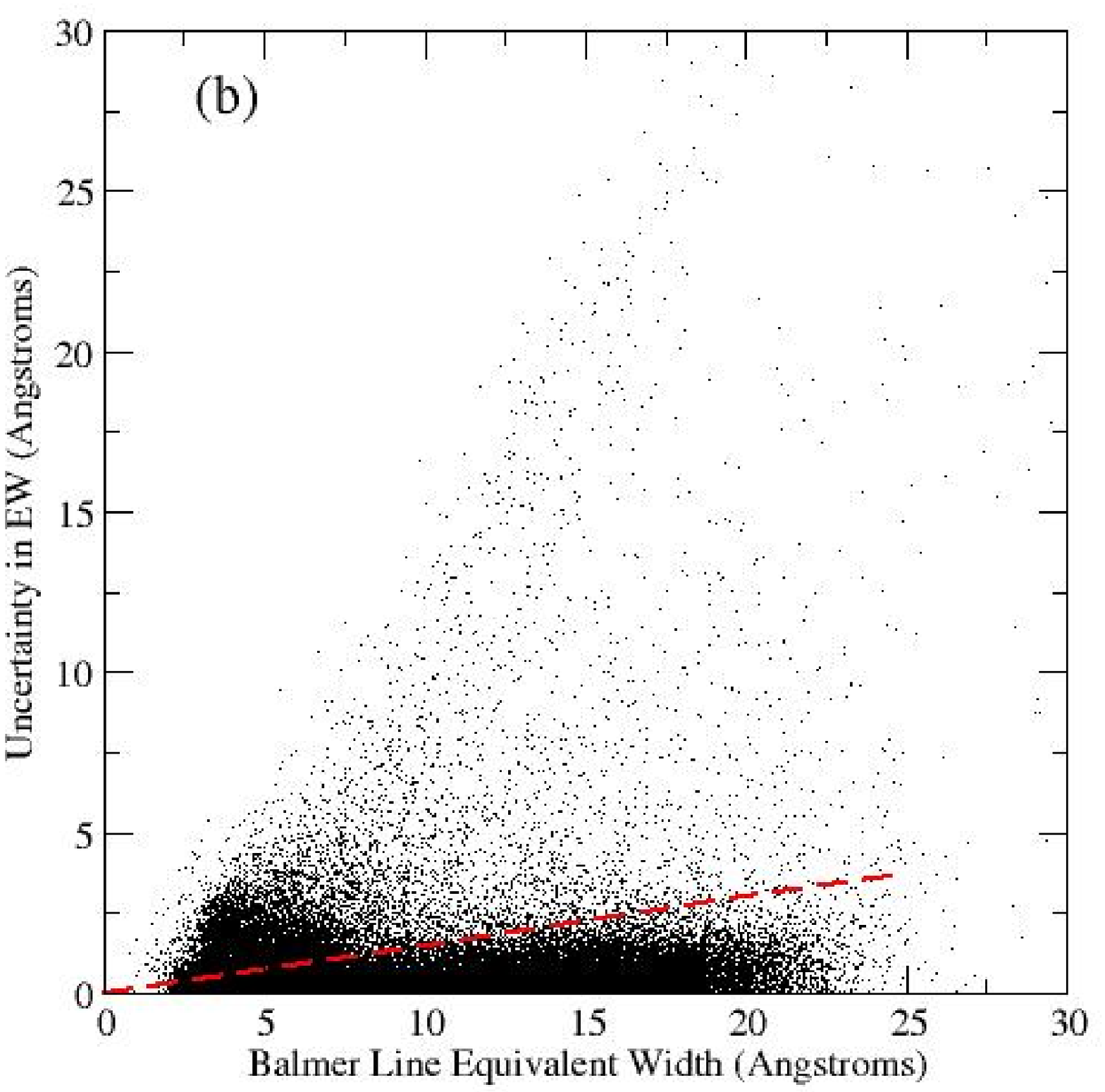}
\caption{SDSS DR-6 sample showing the line-width uncertainty versus Balmer- 
line width for D$_{0.2}$ (a) and EW (b).  The red dashed line is 
a constant $10\%$ limit for D$_{0.2}$ and a $15\%$ limit for EW, which is used to 
remove excessively uncertain line widths.
\label{fig2}}
\end{figure}


\begin{figure}
\epsscale{1.0}
\plotone{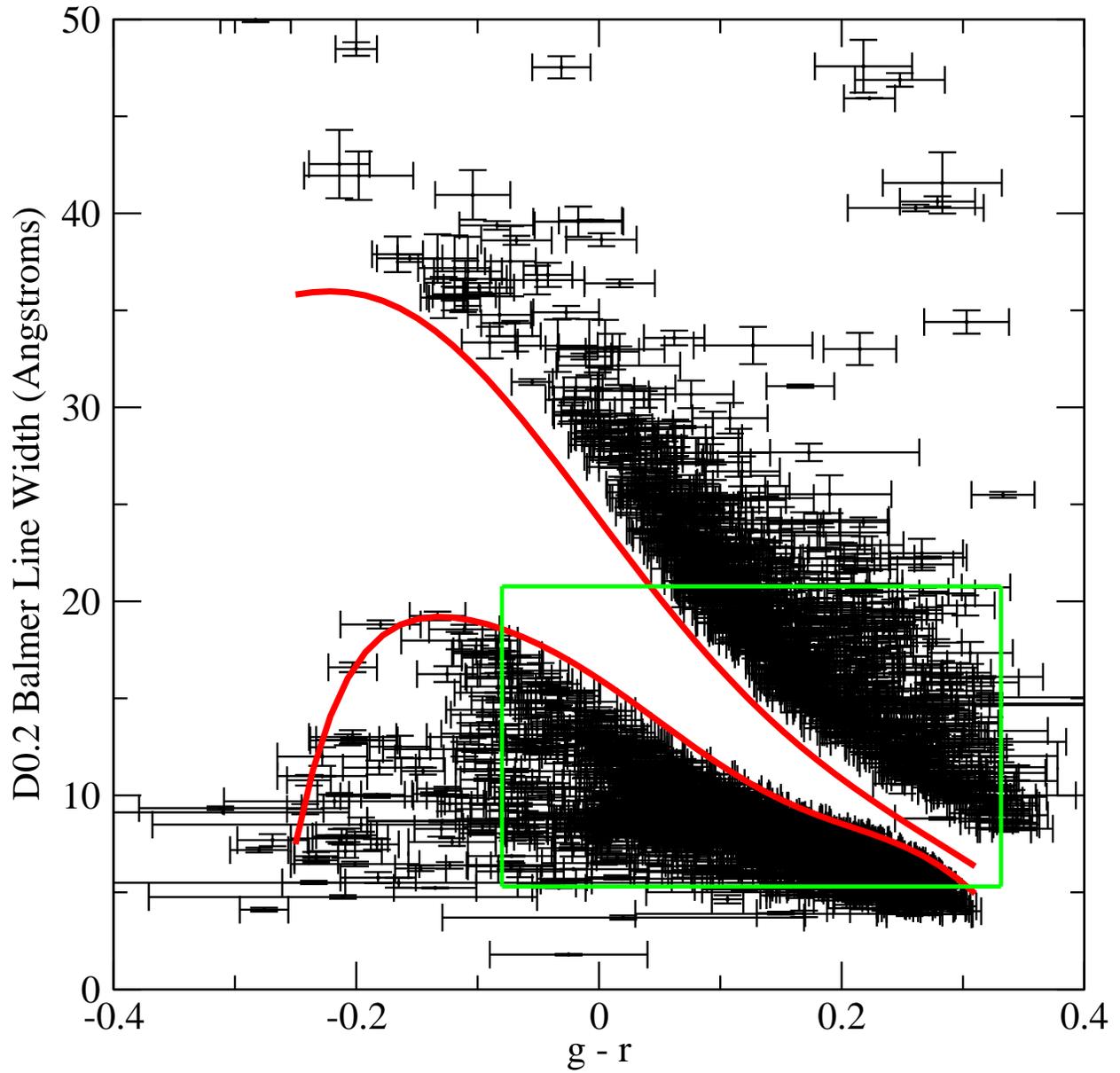}
\caption{D$_{0.2}$ as a function of $g-r$ color for data that we identify
as likely variables, using a D$_{0.2}$ uncertainty limit set at $5\%$. The green box
represents the theoretical limits for instability strip parameters.
\label{fig3}}
\end{figure}

\begin{figure}
\epsscale{1.0}
\plotone{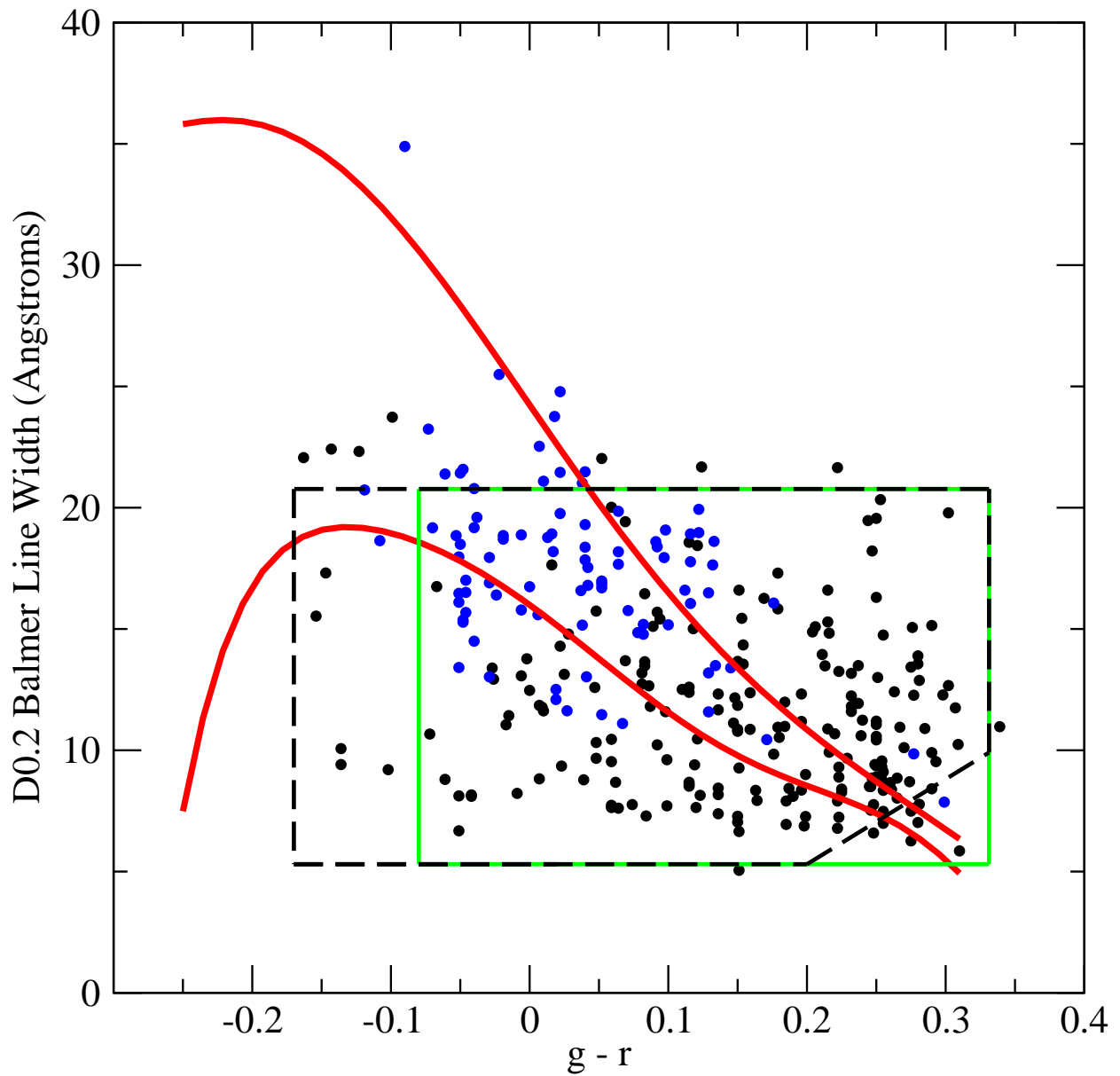}
\caption{D$_{0.2}$ as a function of $g-r$ color for LMCC variables with 
available spectroscopy from SDSS Stripe 82.  
The black dots are RRab variables, while the blue dots are RRc variables. The
black dashed box is the adopted, empirical, instability strip parameter limits. 
The green bounding box is the theorectical limits as in Figure 3.  The majority
of the LMCC sample falls within the confines of the empirical limits.
\label{fig4}}
\end{figure}


\begin{figure}
\epsscale{1.0}
\plottwo{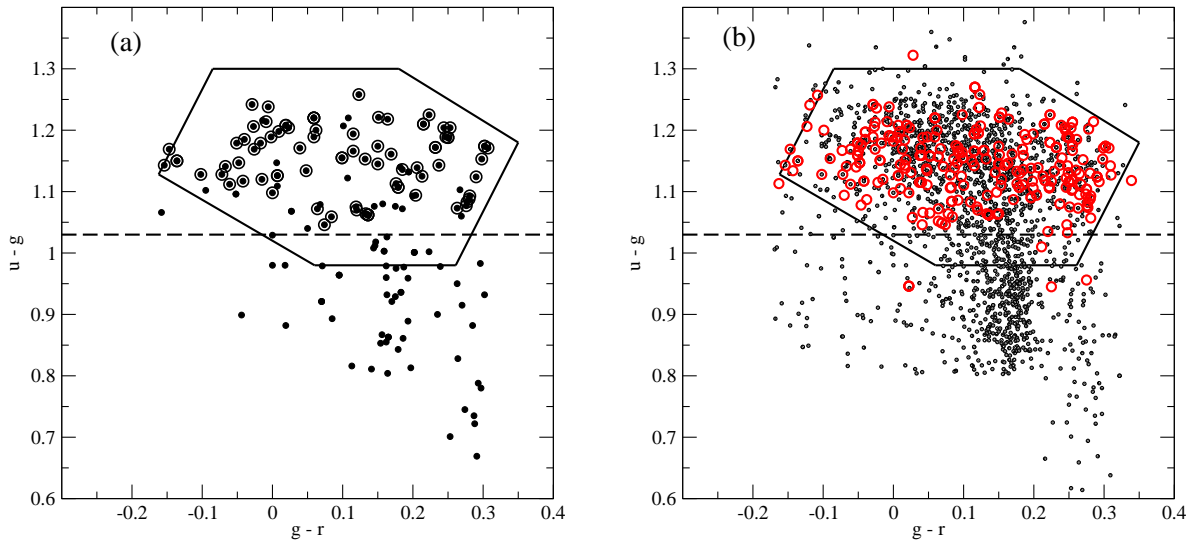}{RWfig5b.eps}
\caption{Color-color plot of the Stripe 82 sample, using the 10/15 uncertainty
criteria described in the text (a). The small dots represent
predicted variables from this technique, while the larger circles are known RR
Lyrae variables from the LMCC. The hexigonal box is based on the color criteria by
Ivezi\'c et al. (2005), while the dashed line is our preferred $u-g$ color cut for variable stars.
Panel (b) shows the entire LMCC spectroscopic sample (in red) and our
entire 10/15 sample (in black). Our chosen $u-g$ color cut suggests a large gain
in efficiency with mimimal loss in completeness.
\label{fig5}}
\end{figure}


\begin{figure}
\epsscale{1.0}
\plotone{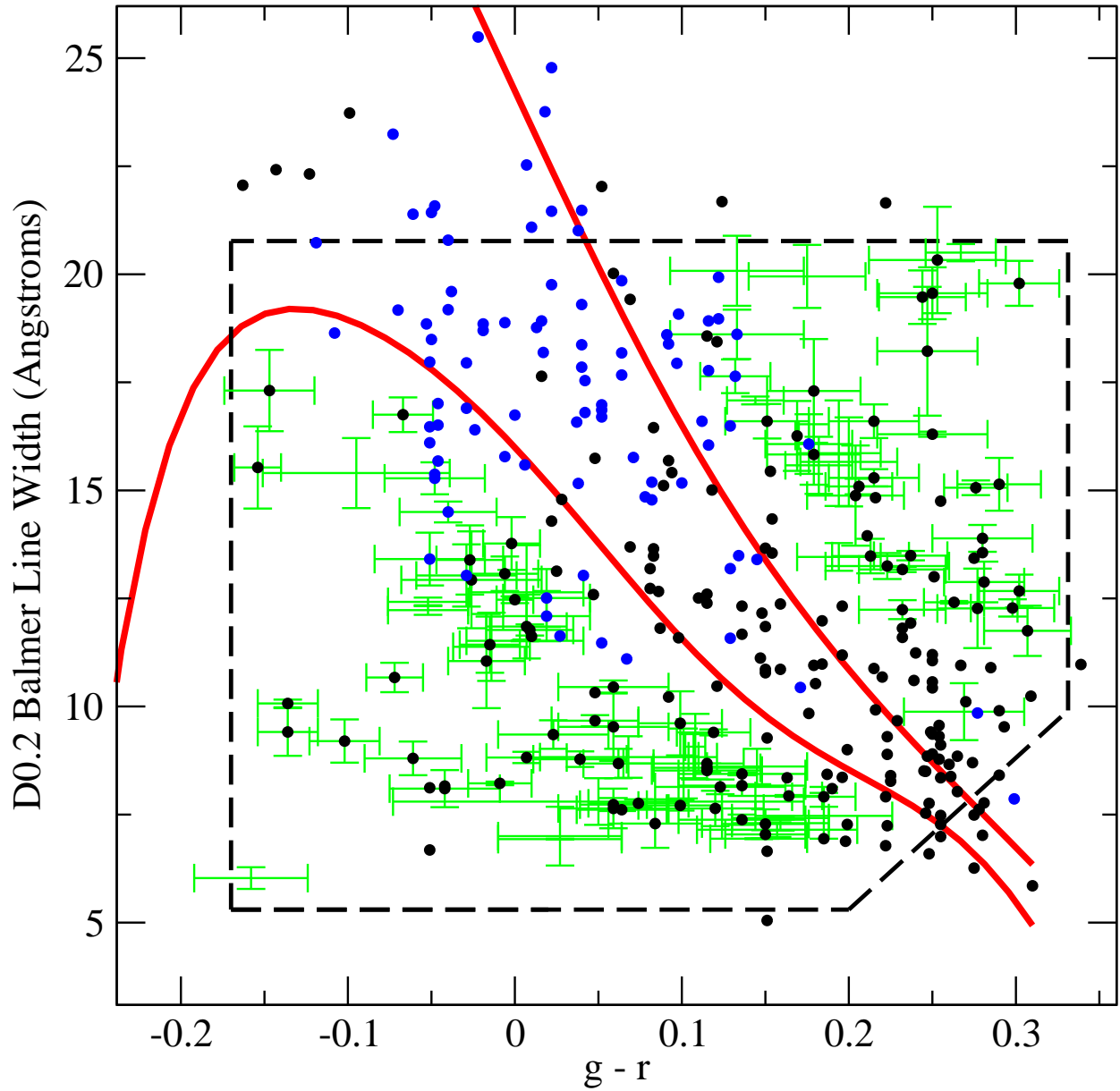}
\caption{D$_{0.2}$ as a function of $g-r$ color for our final DR-6
sample with an imposed color cut.  The dots are as defined in Figure 4. 
The green error bars are the predicted RR Lyrae variables from our technique.
\label{fig6}}
\end{figure}



\begin{figure}
\epsscale{1.0}
\plottwo{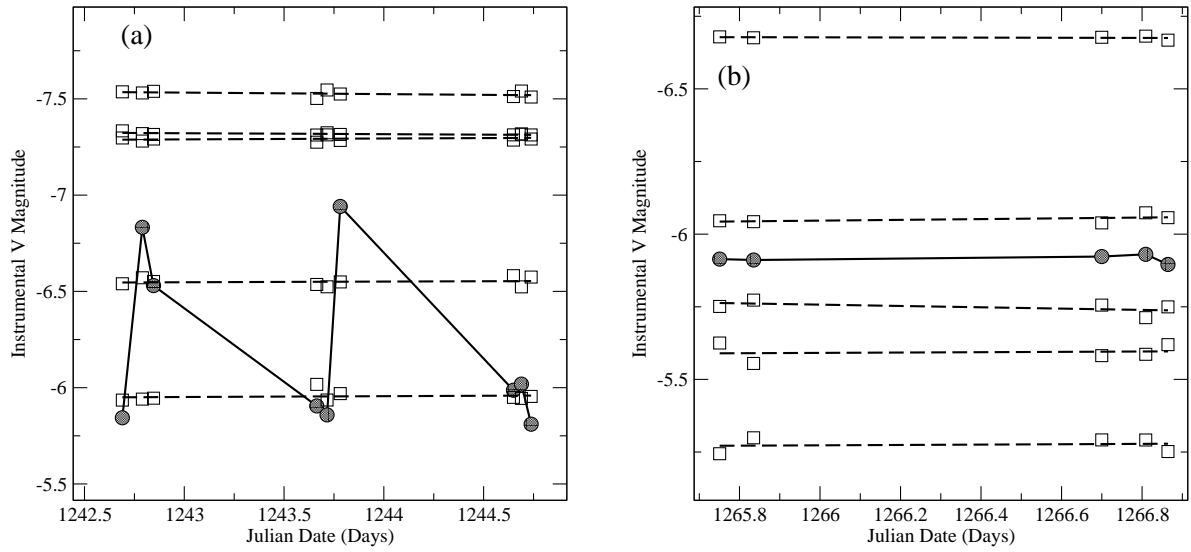}{RWfig7c.eps}
\caption{(a) Light curve (instrumental $V$ magnitudes) for candidate SDSS J130537.39+595957.6, showing strong 
indication of variability, shown as black dots. (b) The flat light
cuve for candidate SDSS J171850.00+264608.0, indicating non-variability. The open squares
in both plots are observations of other non-variable stars in each field.
\label{fig7}}
\end{figure}
 

\begin{figure}
\epsscale{1.0}
\plotone{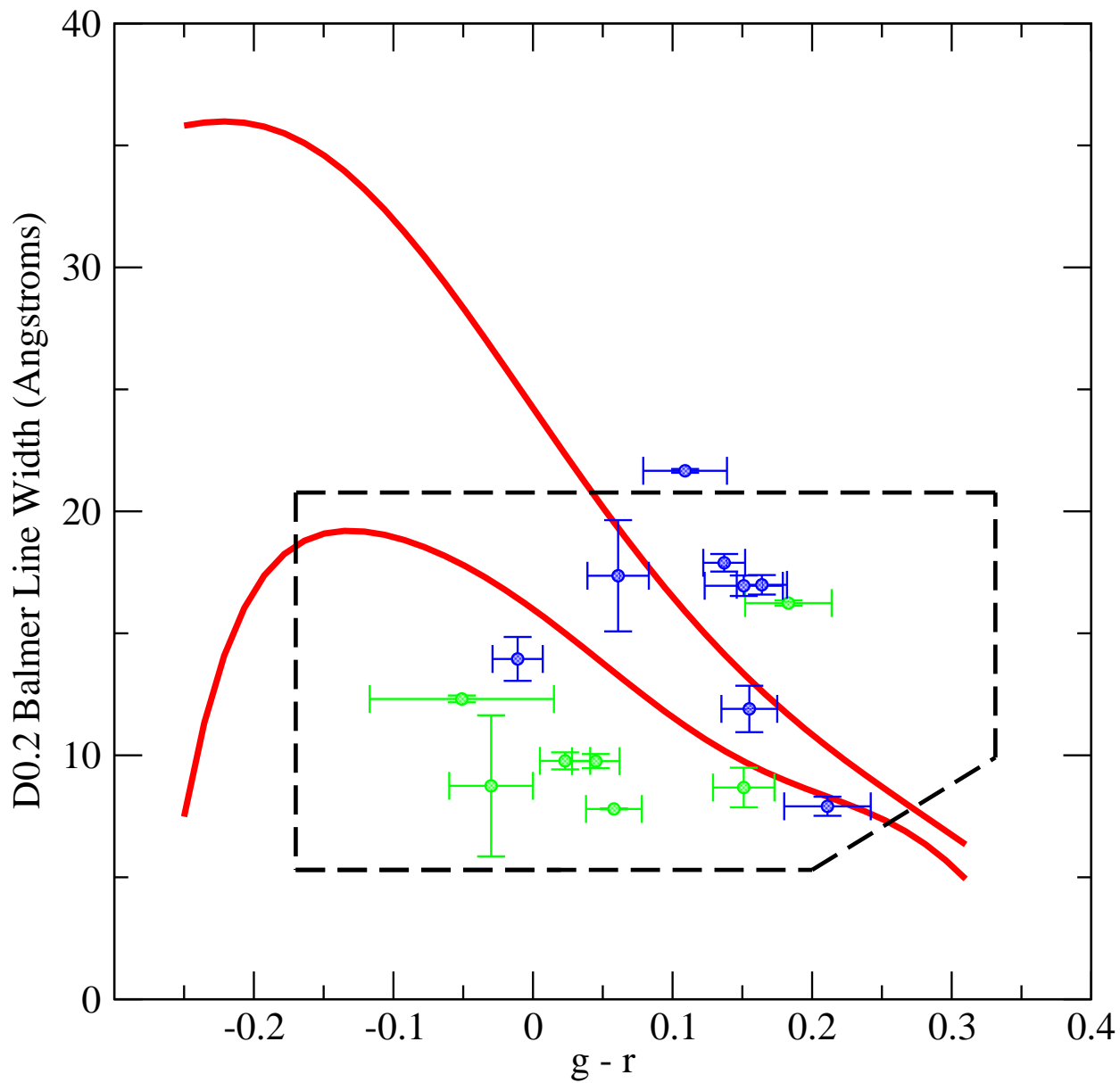}
\caption{D$_{0.2}$ as a function of $g-r$ color for 15 our of variable candidates
with data from the McDonald Observatory. The blue symbols 
indicate non-variability, while the green symbols are stars for
which variability has been detected.  The bounding box is the 
emipirical limits for instability as shown in Figure 4. 
\label{fig8}}
\end{figure}
 
\begin{figure}
\epsscale{1.0}
\plottwo{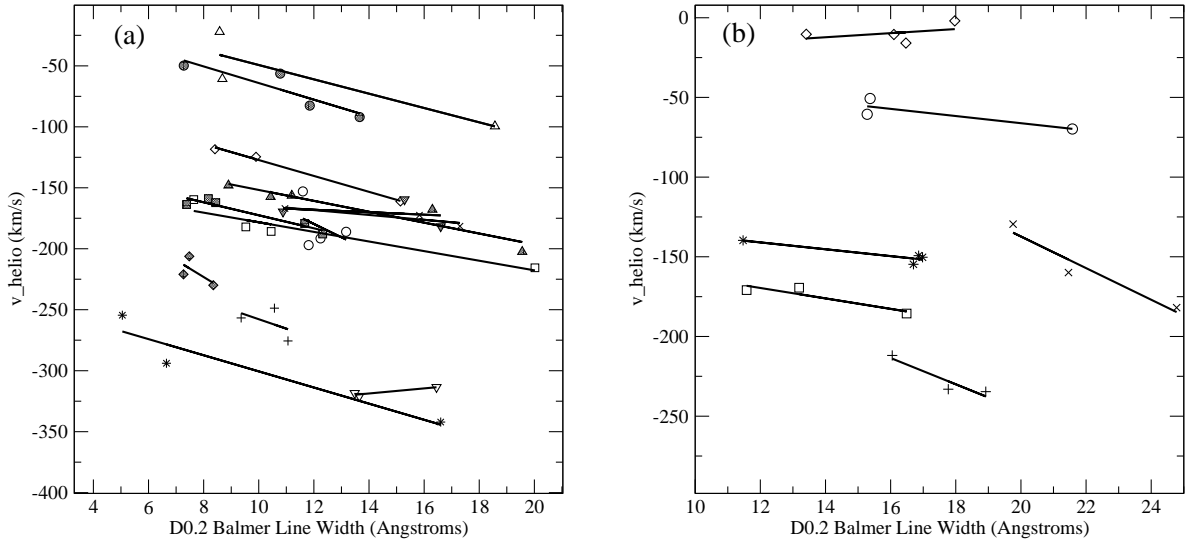}{RWfig9b.eps}
\caption{Heliocentric radial velocity as a 
function of D$_{0.2}$ for the LMCC RRab sample (a) and the 
LMCC RRc sample (b) having with $N \ge 3$ spectroscopic
observations.  Solid lines are linear fits to each star's parameters.
The expected anti-correlation between line-width and radial velocity 
is clearly seen in the RRab sample, but less evident in the RRc sample.
\label{fig9}}
\end{figure}


 
\begin{figure}
\epsscale{1.0}
\plotone{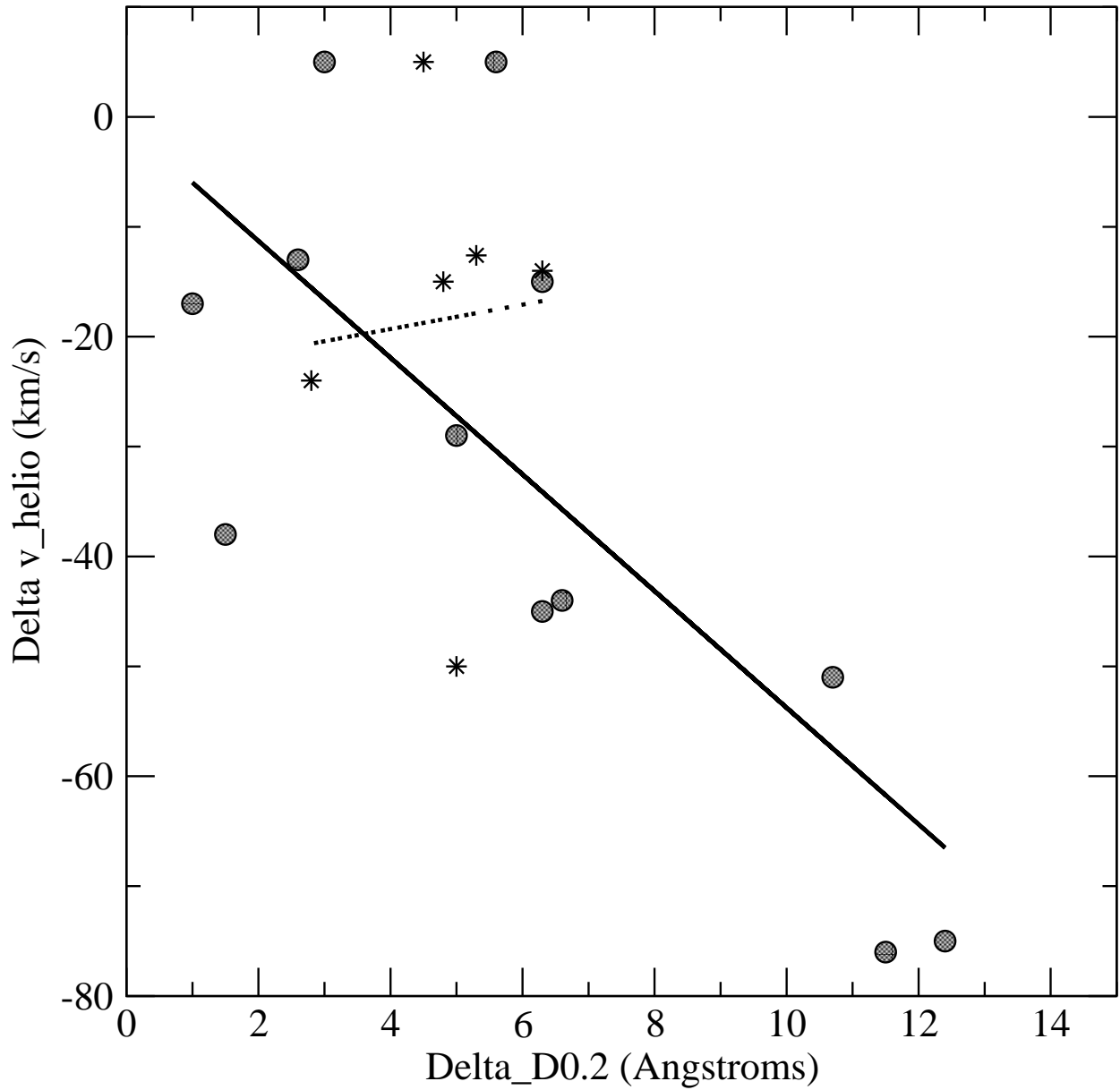}
\caption{Change in heliocentric radial velocity as a function of the change in 
D$_{0.2}$ for the RRab sample (black dots) and the RRc sample (stars) from
Figure 9.
The RRab sample exhibits a range in velocity that is consistent 
with expectations for RRab variables.  The RRc sample shows a 
much smaller change in velocity.
\label{fig10}}
\end{figure}


\begin{figure}
\epsscale{1.0}
\plotone{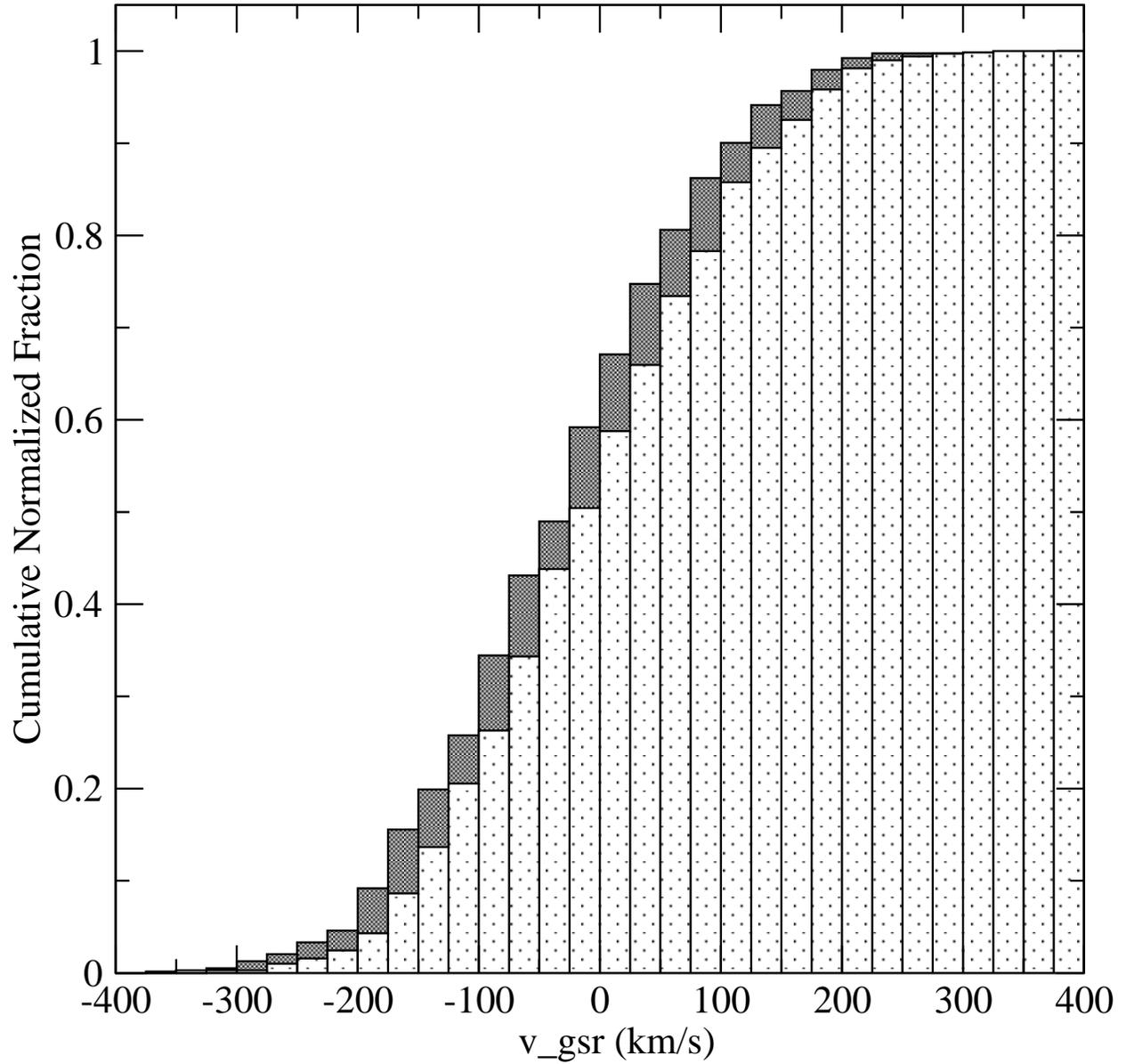}
\caption{Normalized cumulative distribution plot of $v_{gsr}$ for the 
sample with spectral lines near maximum (filled) and near minimum (unfilled)
light.  There is an overall shift between the two distributions, with 
maximum light being more negative.
\label{fig11}}
\end{figure}

\begin{figure}
\epsscale{1.0}
\plotone{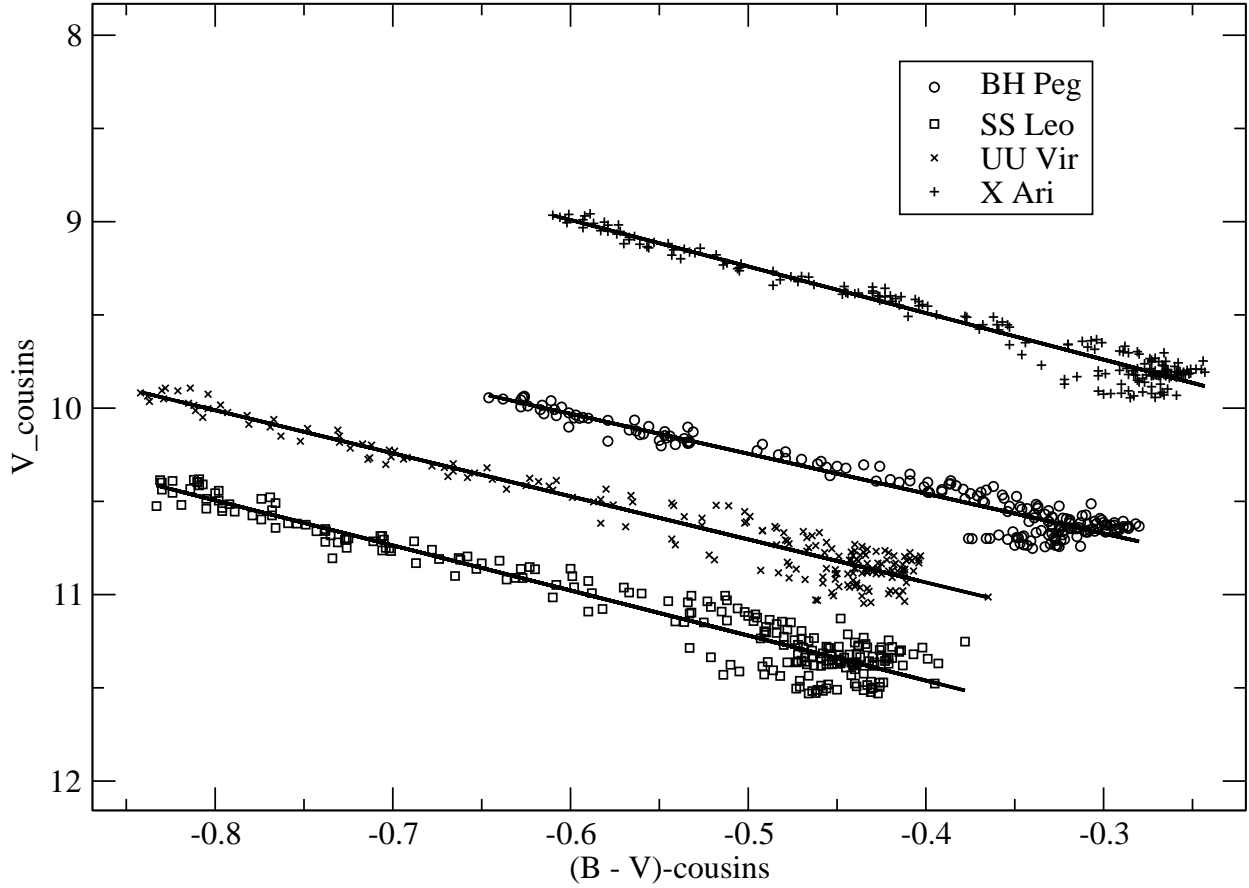}
\caption{Cousins $V$ magnitude as a function of $B-V$ color from Carrillo et al.
(1995)
for four well-studied RR Lyrae variables.  Strong correlations are evident
between apparent magnitude and color over the entire phase of each star. 
\label{fig12}}
\end{figure}

\begin{figure}
\epsscale{1.0}
\plottwo{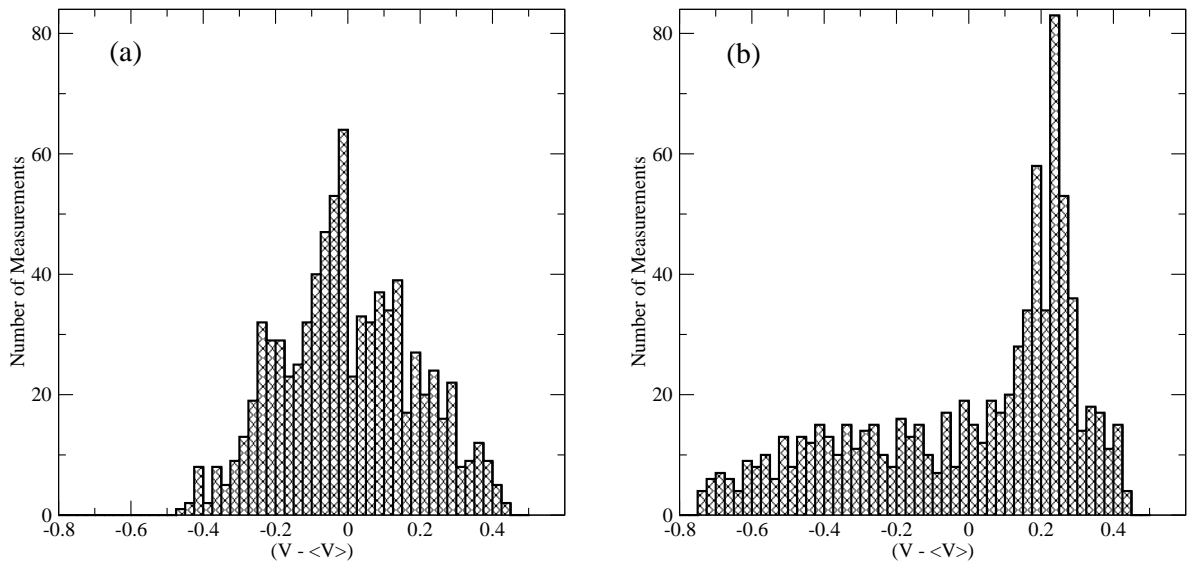}{RWfig13b.eps}
\caption{(a) Distribution of the difference between predicted
$V$ magnitude and average $V$ magnitude for stars from Carrillo et al.
(1995). (b)
Difference between the measured $V$ and $<V>$.  As is clear, 
distribution (b) shows much more scatter than the predicted $V$ magnitudes.
\label{fig13}}
\end{figure}



\begin{figure}
\epsscale{1.0}
\plottwo{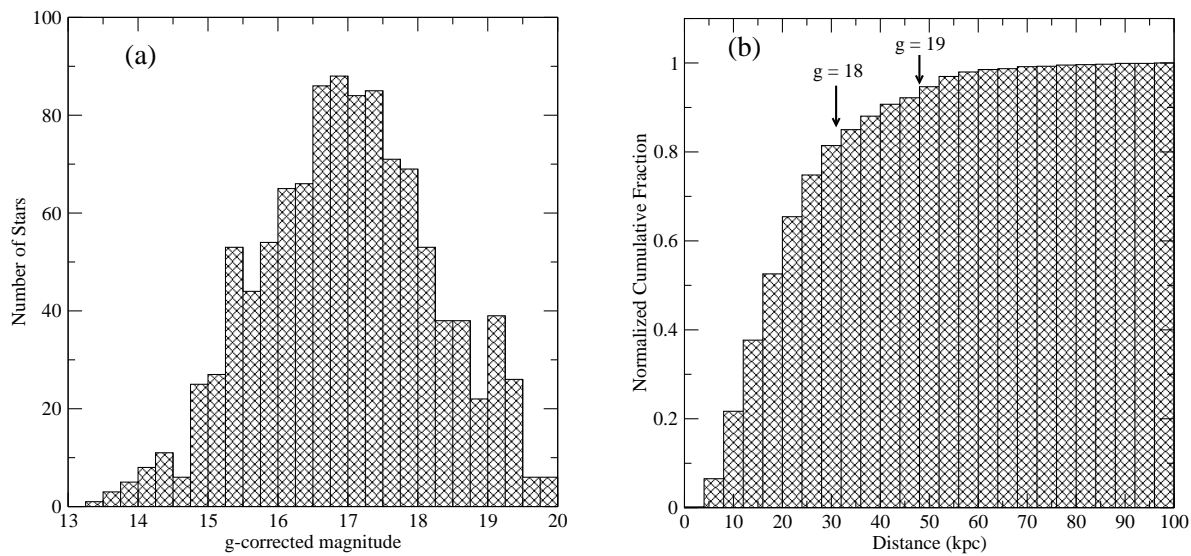}{RWfig14b.eps}
\caption{(a) Distribution of 
predicted $<g>$ magnitudes for the SDSS DR-6 sample.  There is a rapid
decline in the number of stars beyond $<g> = 18$, for reasons described in the
text.  (b) Normalized cumulative distribution of the estimated
distances for this same sample. The positions of the predicted $<g> = $18 and
$<g> = $19 are shown. We expect high efficiency rates for $94\%$ of our sample,
out to a distance of $\sim48$ kpc.
\label{fig14}}
\end{figure}



\begin{figure}
\epsscale{1.0}
\plottwo{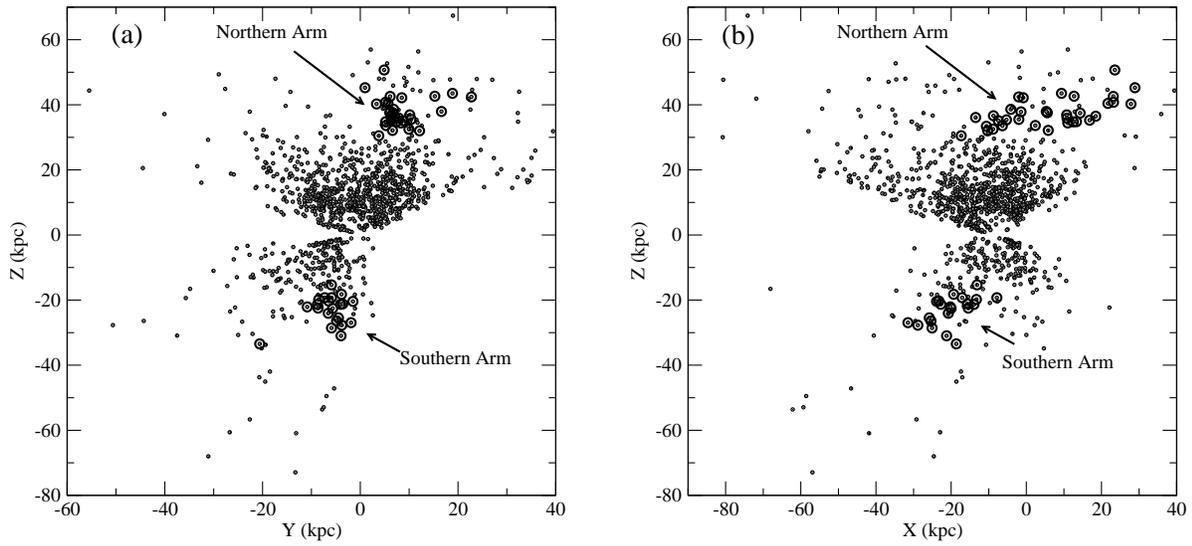}{RWfig15b.eps}
\caption{Plots of the Galactic Y-Z (a) and X-Z (b) projection for the DR-6 RR Lyrae sample.  
The outer halo appears to be decidedly non-homogeneous, although this is
influenced by the choice of targets for spectroscopy.
Stars expected to be members of the Sagittarius Stream are shown as large
circles. The Northern Arm appears very ``stream-like'' in the X-Z projection.
\label{fig15}}
\end{figure}


\begin{figure}
\epsscale{1.0}
\plotone{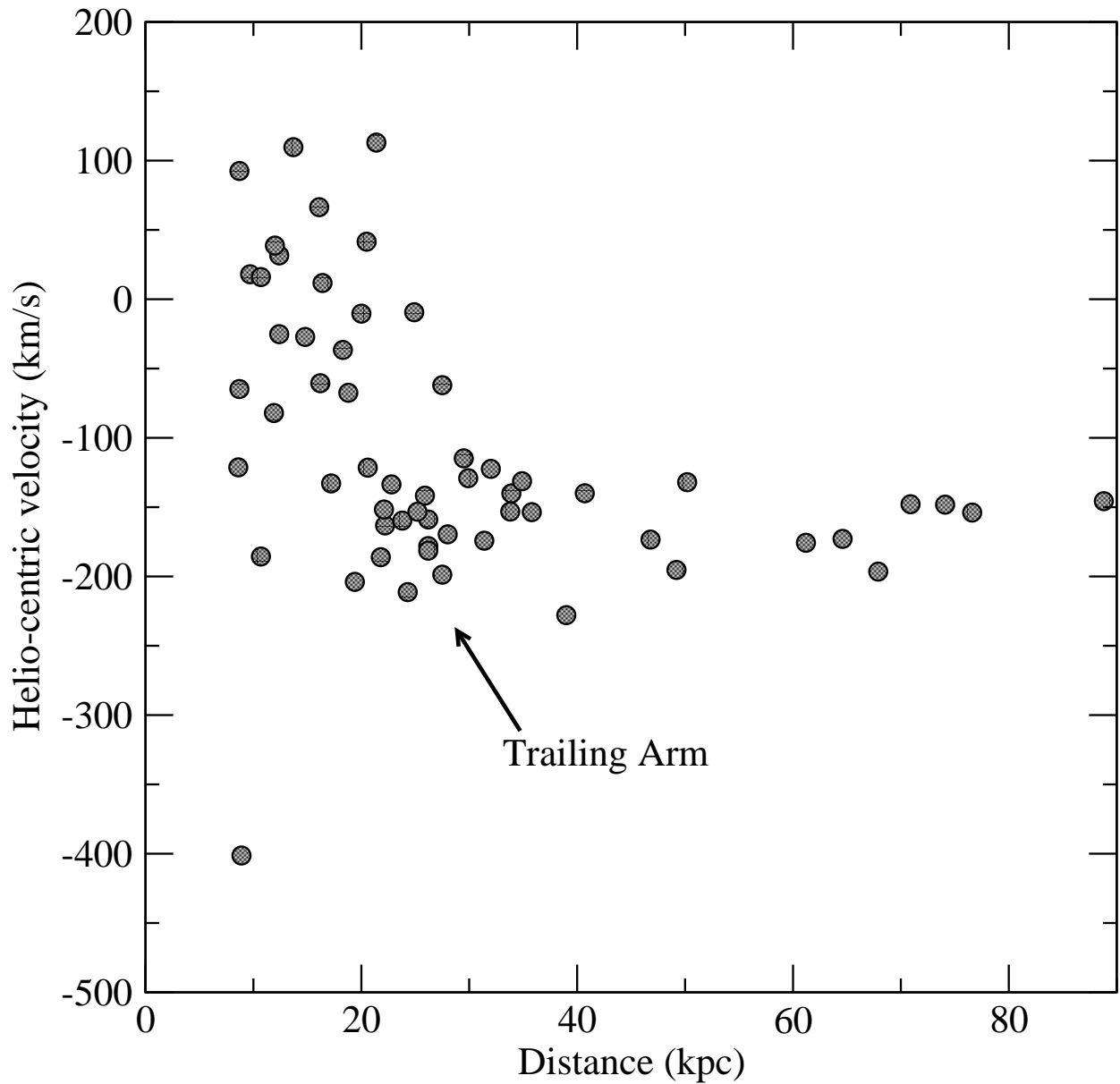}
\caption{Heliocentric radial velocity as a 
function of distance for likely members of the trailing Southern Sagittarius
Stream.  The overdensity of stars in the plot is a positive detection of the stream.
\label{fig16}}
\end{figure}

\begin{figure}
\epsscale{1.0}
\plotone{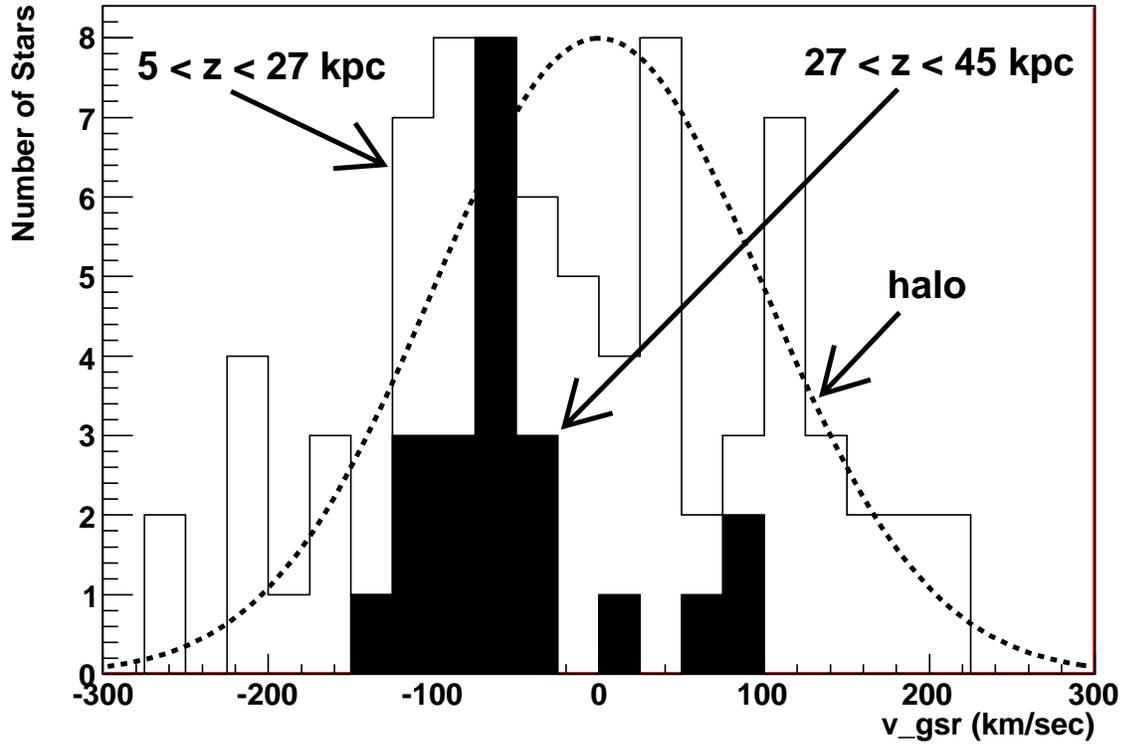}
\caption{Histograms of the $v_{gsr}$ near the NPC for samples split in 
distance above the 
Galactic plane, Z.  The more distant sample is dominated by negative-velocity
stars, while the more nearby sample appears to have structure that
is inconsistent with sampling from the halo field population (shown as a
superposed Gauassian with mean velocity $v_{gsr} = 0$ km $s^{-1}$, and dispersion
100 km $s^{-1}$).
\label{fig17}}
\end{figure}

\clearpage





\begin{deluxetable}{lccccc}
\tabletypesize{\scriptsize}
\tablecaption{Model Boundaries for Variability \label{tbl-1}}
\tablewidth{0pt}
\tablehead{
\colhead{Coefficient} & \colhead{Upper Bound D$_{0.2}$} &
\colhead{Lower Bound D$_{0.2}$} &
\colhead{Upper Bound Eq. Width} & \colhead{Lower Bound Eq. Width}
}
\startdata
a0 & 24.242 & 15.995 & 16.316 & 11.256  \\
a1 & -83.185 & -41.292 & -49.251 & -24.542  \\
a2 & 8.5474  & -85.6  & -6.1093 & -16.395  \\
a3 & 602.91 & 436.57 & 383.51 & -47.15  \\
a4 & -980.24 & 1578.5 & -850.23 & -254.08  \\
a5 & -2079.6 & -556.16 & -1290  & 5854.4  \\
a6 & 5379.8  & -38376 & 6758.8  & -17999  \\
a7 & -2410.3  & 65538 & -6108.3 & 16391  \\
\enddata
\end{deluxetable}

\clearpage

\begin{deluxetable}{lrrcrc}
\tabletypesize{\scriptsize}
\tablecaption{Completeness for Stripe82 RR Lyrae Variables \label{tbl-2}}
\tablewidth{0pt}
\tablehead{
\colhead{Samples} & \colhead{S82 Sample} & \colhead{Recovered ($5/10$)} &
\colhead{Percentage ($5/10$)} & \colhead{Recovered ($10/15$)} &
\colhead{Percentage ($10/15$)}
}
\startdata
Total Sample & 298 & 64 & 21.5 & 86 & 28.9 \\
Inside Instability Limits & 266 & 60 & 22.5 & 82 & 30.8 \\
Total In-phase & 152 & $-$ & $-$ & $-$  & $-$  \\
In-phase Inside Instability Limits & 144 & $-$ & $-$ & $-$  & $-$  \\
Out-phase Inside Instability Limits & 142 & 60 & 42.2 & 82 & 57.7 \\
Out-phase Type RRab & 128 & 53 & 41.4 & 73 & 57.0 \\
Out-phase Type RRc & 14 & 7 & 50.0 & 9 & 64.3 \\
\enddata
\end{deluxetable}

\clearpage

\begin{deluxetable}{lrrcrrc}
\tabletypesize{\scriptsize}
\rotate
\tablecaption{Percentage of Correct Identifications \label{tbl-3}}
\tablewidth{0pt}
\tablehead{
\colhead{Samples} & \colhead{Total ID ($5/10$)} & \colhead{Correct ID ($5/10$)} &
\colhead{Percentage($5/10$)} & \colhead{Total ID ($10/15$)} &
\colhead{Correct ID ($10/5$)} & \colhead{Percentage ($10/15$)}
}
\startdata
Total Variable Sample & 225 & 64 & 28.4 & 355 & 86 & 24.2 \\
Inside Instability Limits & 111 & 60 & 54.1 & 168 & 82 & 48.8 \\
$g < 19.0$  & 95 & 60 & 63.2 & 137 & 82 & 59.8 \\
$g < 18.0$  & 66 & 48 & 72.7 & 100 &  69 & 69.0 \\
Including color cut & 71 & 60 & 84.5 & 107 & 82 & 76.6 \\
$g < 19.0$  & 68 & 60 & 88.2 & 98 & 82 & 83.7 \\
$g < 18.0$  & 53 & 48 & 90.6 & 82 & 69 & 84.1 \\

\enddata
\end{deluxetable}

\clearpage

\begin{deluxetable}{lrrcrrc}
\tabletypesize{\scriptsize}
\rotate
\tablecaption{Percentage of Correct Identifications for Various Samples  \label{tbl-4}}
\tablewidth{0pt}
\tablehead{
\colhead{Samples} & \colhead{Total ID ($5/10$)} & \colhead{Correct ID ($5/10$)} &
\colhead{Percentage ($5/10$)} & \colhead{Total ID ($10/15$)} &
\colhead{Correct ID ($10/5$)} & \colhead{Percentage ($10/15$)}
}
\startdata
Single Epoch Spec & 45 & 34 & 75.5 & 71 & 47 & 66.2 \\
$g < 19.0$  & 42 & 34 & 80.9 & 62 & 47 & 75.8 \\
$g < 18.0$  & 35 & 30 & 85.7 & 54 & 42 & 77.7 \\
Multiples Counted Once  & 65 & 54 & 83.1 & 92 & 68 & 73.9 \\
$g < 19.0$  & 62 & 54 & 87.1 & 83 & 68 & 81.9 \\
$g < 18.0$  & 50 & 45 & 90.0 & 70 & 59 & 84.3 \\
A

\enddata
\end{deluxetable}

\clearpage
\begin{deluxetable}{lccccc}
\tabletypesize{\scriptsize}
\tablecaption{McDonald Results -- February 2007 \label{tbl-5}}
\tablewidth{0pt}
\tablehead{
\colhead{Name} & \colhead{n} & \colhead{$\xi_{<V>}$} &
\colhead{$\sigma_{V}$} & \colhead{$\chi{^2}_{V}$} &
\colhead{Variability}
}
\startdata

SDSS J100659.53+533318.9 & 8 & 0.025 & 0.019 & 0.055 & No \\
SDSS J100719.58+532845.1 & 8 & 0.019 & 0.258 & 3.763 & Yes \\
SDSS J095317.79+002451.3 & 8 & 0.032 & 0.167 & 0.700 & Maybe \\
SDSS J095055.49+003253.9 & 8 & 0.080 & 0.130 & 0.020 & No \\
SDSS J134427.48+002410.9 & 11 & 0.134 & 0.000 & 0.074 & No \\
SDSS J134319.13-000622.0 & 11 & 0.020 & 0.093 & 0.566 & Maybe \\

\enddata
\end{deluxetable}

\clearpage
\begin{deluxetable}{lccccccccc}
\tabletypesize{\scriptsize}
\tablecaption{McDonald Results -- May 2007 \label{tbl-6}}
\tablewidth{0pt}
\tablehead{
\colhead{Name} & \colhead{$n_{V}$} & \colhead{$n_{R}$} &
\colhead{$\xi_{<V>}$} & \colhead{$\xi_{<R>}$} &
\colhead{$\sigma_{V}$} & \colhead{$\sigma_{R}$} &
\colhead{$\chi{^2}_{V}$} & \colhead{$\chi{^2}_{R}$} &
\colhead{Variability}
}
\startdata

SDSS J132158.02+290807.0  & 9 & 9 & 0.023 & 0.047 & 0.179 & 0.146 & 1.61 & 1.072 & Maybe \\
SDSS J130141.75+515158.3  & 10 & 9 & 0.029 & 0.024 & 0.087 & 0.109 & 0.313 & 0.508 & No \\
SDSS J130537.39+595957.6  & 9 & 9 & 0.0180 & 0.014 & 0.453 & 0.369 & 19.978 & 16.618 & Yes \\
SDSS J133003.97+605104.8  & 8 & 9 & 0.0120 & 0.010 & 0.194 & 0.197 & 3.652 & 4.385 & Yes \\
SDSS J130707.52+580039.2  & 10 & 10 & 0.026  & 0.023 & 0.217 & 0.211 & 2.821 & 2.845 & Yes \\

\enddata
\end{deluxetable}

\clearpage
\begin{deluxetable}{lccccccccc}
\tabletypesize{\scriptsize}
\tablecaption{McDonald Results -- June 2007 \label{tbl-7}}
\tablewidth{0pt}
\tablehead{
\colhead{Name} & \colhead{$n_{V}$} & \colhead{$n_{B}$} &
\colhead{$\xi_{<V>}$} & \colhead{$\xi_{<B>}$} &
\colhead{$\sigma_{V}$} & \colhead{$\sigma_{B}$} &
\colhead{$\chi{^2}_{V}$} & \colhead{$\chi{^2}_{B}$} &
\colhead{Variability}
}
\startdata

SDSS J165340.87+342302.8  & 3 & 3 & 0.010 & 0.022 & 0.159 & 0.227 & 2.641 & 6.064 & Yes \\
SDSS J170013.06+320148.7  & 5 & 5 & 0.019 & 0.021 & 0.298 & 0.474 & 5.204 & 13.19 & Yes \\
SDSS J170545.77+201036.7  & 5 & 5 & 0.055 & 0.047 & 0.187 & 0.276 & 0.818 & 1.379 & Maybe \\
SDSS J171850.00+264608.0  & 5 & 5 & 0.019 & 0.012 & 0.012 & 0.045 & 0.032 & 0.210 & No \\
SDSS J171909.57+292027.8  & 4 & 4 & 0.021 & 0.026 & 0.516 & 0.664 & 20.85 & 35.840 & Yes \\
SDSS J172300.39+274401.5  & 7 & 5 & 0.038 & 0.004 & 0.031 & 0.062 & 0.069 & 0.150 & No \\
SDSS J223230.81-082856.3  & 5 & 5 & 0.031 & 0.075 & 0.153 & 0.000 & 0.875 & 0.078 & No \\
SDSS J223331.14-084159.2  & 5 & 5 & 0.061 & 0.037 & 0.076 & 0.037 & 0.147 & 1.123 & No \\

\enddata
\end{deluxetable}

\clearpage

\begin{deluxetable}{llrrrrrrrrrrrcc}
\tabletypesize{\scriptsize}
\rotate
\tablecaption{The RR Lyrae Sample \label{tbl-8}}
\tablewidth{0pt}
\tablehead{
\colhead{Name} & \colhead{Spec} & \colhead{$\alpha$} & \colhead{$\delta$} &
\colhead{l} & \colhead{b} &
\colhead{CaIIK EW} & \colhead{$<H EW>$} & 
\colhead{$<H D_{0.2}>$} & 
\colhead{g$_{cor}$} & \colhead{(u-g)$_{o}$} & \colhead{(g-r)$_{o}$} &
\colhead{v$_{helio}$} & \colhead{N} & \colhead{Sample}
}
\startdata

SDSS J094322.02-001640.0 & 51602-0266-182 & 145.841766 &   -0.277789 &  236.2 &   
37.2 &  2.34 & 12.81 &  19.44 & 16.862 & 1.113 &  0.145 &  249.9 &      &   5$/$10 \\
SDSS J100924.16+004923.5 & 51909-0270-403 & 152.350662 &    0.823192 &  240.1 &   
43.1 &  1.30 & 13.76 &  20.43 & 16.797 & 1.229 &  0.091 &  170.8 &    2 &   5$/$10 \\
SDSS J103243.31+010231.2 & 51957-0273-523 & 158.180466 &    1.041989 &  245.2 &   
47.7 &  2.29 & 10.97 &  16.60 & 15.279 & 1.176 & -0.077 &  -73.4 &      &       \\
SDSS J104604.62+003617.7 & 51910-0275-493 & 161.519241 &    0.604916 &  249.1 &   
49.9 &  2.00 &  8.65 &  12.97 & 14.392 & 1.248 &  0.016 &  215.3 &      &       \\
SDSS J112425.37-000919.7 & 51612-0280-101 & 171.105713 &   -0.155471 &  261.9 &   
55.6 &  3.76 &  5.16 &   8.16 & 17.713 & 1.118 &  0.042 &  166.3 &      &       \\
SDSS J113335.65-011012.8 & 51630-0282-203 & 173.398529 &   -1.170210 &  266.3 &   
56.1 &  4.06 & 11.22 &  17.10 & 16.553 & 1.263 & -0.080 &    1.0 &    2 &   5$/$10 \\
SDSS J113005.74+001107.8 & 51658-0282-343 & 172.523926 &    0.185493 &  263.6 &   
56.7 &  3.10 &  5.43 &   7.79 & 17.653 & 1.080 &  0.183 &  138.4 &      &   5$/$10 \\
SDSS J113821.89+004508.3 & 51630-0282-608 & 174.591217 &    0.752307 &  266.3 &   
58.3 &  6.84 &  4.70 &   6.10 & 17.683 & 1.042 &  0.001 &   95.7 &    2 &       \\
SDSS J120910.44-004759.1 & 52023-0287-296 & 182.293488 &   -0.799740 &  281.2 &   
60.3 &  3.33 &  5.28 &   8.51 & 17.266 & 1.298 & -0.026 &   74.0 &      &   5$/$10 \\
SDSS J121638.40-003711.8 & 52000-0288-279 & 184.160004 &   -0.619936 &  284.7 &   
61.0 &  1.64 & 13.00 &  20.33 & 19.020 & 1.061 &  0.136 &  -42.8 &      &       \\

\enddata
\tablecomments{Table \ref{tbl-8} is published in its entirety in the
electronic edition of the {\it Astronomical Journal}.  A portion is
shown here for guidance regarding its form and content.}
\end{deluxetable}

\clearpage

\end{document}